\documentclass{article}
\usepackage[final,nonatbib]{neurips_2021}
\usepackage{microtype}
\usepackage{graphicx}
\usepackage{bmpsize}
\usepackage{booktabs} % for professional tables
\usepackage{amsmath}
\usepackage{amssymb}
\usepackage[colorinlistoftodos,textsize=scriptsize]{todonotes}
\usepackage{marginnote}
\usepackage{amsthm}
\usepackage{booktabs}
\usepackage{multirow}
\usepackage{float}
\usepackage{subcaption}
\usepackage{wrapfig}
\graphicspath{{./Figures/}}
\usepackage{hyperref}
\usepackage[utf8]{inputenc}
\usepackage{bbm}
\usepackage{enumitem}
\usepackage{pifont}
\usepackage{xcolor}
\usepackage[frozencache=true,cachedir=.]{minted}
\newif\ifarxivsubmit
\arxivsubmittrue

\title{Latent Execution for Neural Program Synthesis}
\author{
    Xinyun Chen \\
    UC Berkeley \\ 
    \texttt{xinyun.chen@berkeley.edu} 
    \And 
    Dawn Song \\
    UC Berkeley \\ 
    \texttt{dawnsong@cs.berkeley.edu}
    \And
    Yuandong Tian \\
    Facebook AI Research \\
    \texttt{yuandong@fb.com}
}

\begin{document}
\def\ours{\texttt{LaSynth}}
\def\laet{\texttt{LaET}}
\newcommand{\eat}[1]{}
\newif\ifsubmit
\submittrue

\newcommand{\maybetodo}[1]{\ifsubmit{}\else{#1}\fi}

\ifsubmit
\newcommand{\xinyun}[1]{{\maybetodo{\color{blue}}}}
\newcommand{\yuandong}[1]{{\maybetodo{\color{red}}}}
\else
\newcommand{\xinyun}[1]{{\maybetodo{\color{blue}[Xinyun: #1]}}}
\newcommand{\yuandong}[1]{{\maybetodo{\color{red}[Yuandong: #1]}}}
\fi

\newcommand{\cmark}{\textcolor{green!80!black}{\ding{51}}}
\newcommand{\xmark}{\textcolor{red}{\ding{55}}}

\maketitle

\begin{abstract}
Program synthesis from input-output (IO) examples has been a long-standing challenge. While recent works demonstrated limited success on domain-specific languages (DSL), it remains highly challenging to apply them to real-world programming languages, such as C. Due to complicated syntax and token variation, there are three major challenges: \textbf{(1)} unlike many DSLs, programs in languages like C need to compile first and are not executed via interpreters; \textbf{(2)} the program search space grows exponentially when the syntax and semantics of the programming language become more complex; and \textbf{(3)} collecting a large-scale dataset of real-world programs is non-trivial. As a first step to address these challenges, we propose {\ours} and show its efficacy in a \emph{restricted-C} domain (i.e., C code with tens of tokens, with sequential, branching, loop and simple arithmetic operations but no library call). More specifically, \ours{} learns the latent representation to approximate the execution of partially generated programs, even if they are incomplete in syntax (addressing \textbf{(1)}). The learned execution significantly improves the performance of next token prediction over existing approaches, facilitating search (addressing \textbf{(2)}). Finally, once trained with randomly generated ground-truth programs and their IO pairs, \ours{} can synthesize more concise programs that resemble human-written code. Furthermore, retraining our model with these synthesized programs yields better performance with fewer samples for both Karel and C program synthesis, indicating the promise of leveraging the learned program synthesizer to improve the dataset quality for input-output program synthesis (addressing \textbf{(3)}). When evaluating on whether the program execution outputs match the IO pairs, \ours{} achieves 55.2\% accuracy on generating simple C code with tens of tokens including loops and branches, outperforming existing approaches without executors by around 20\%.~\footnote{The code is available at \url{https://github.com/Jungyhuk/latent-execution}.}
\end{abstract}

\section{Introduction}
\vspace{-0.1in}
Program synthesis from input-output (IO) pairs, also called programming by example (PBE), requires high-level reasoning and remains a challenging problem for deep models. Unlike Natural Language Processing (NLP)~\cite{bahdanau2014neural,devlin2018bert} and perceptual tasks such as Computer Vision (CV)~\cite{deng2009imagenet,he2016deep}, the mapping from IO pairs to the program itself is hard to model. Many works attempt to learn a direct mapping from training samples, but often found that it is already difficult to achieve a low training error, and generalization to new problems is even harder. Alternatively, one might choose to formulate program synthesis as a search problem: to find the program that satisfies IO pairs. Unfortunately, the search space of programs is often vast and highly non-smooth, i.e., a small perturbation of the program often leads to a complete change of the output.  

While there are many previous works on programming by example tasks~\cite{balog2016deepcoder,devlin2017robustfill,bunel2018leveraging}, they mainly focus on Domain Specific Languages (DSLs), and cannot be easily applied to popular general-purpose programming languages. For example, to synthesize C programs, we need to deal with both high-level control flows (e.g., branching and loop) and low-level operations (e.g., which variable is the target of assignment). Moreover, unlike DSLs (e.g., Karel) for which it is feasible to implement a per-line interpreter, C programs need compilation and a partial C program cannot execute. On the other hand, some recent works investigate natural language descriptions as the auxiliary information of the program specification, and they evaluate neural program synthesis models on constrained or simplified competitive programming problems~\cite{kulal2019spoc,alet2021large,hendrycks2021measuring,chen2021evaluating,austin2021program}. Although some of these works demonstrate promising results for synthesizing Python or C code, they require manual annotations of natural language specifications~\cite{kulal2019spoc} or large-scale pre-training on human-written programs~\cite{chen2021evaluating,austin2021program}, and the performance significantly degrades when only input-output examples are fed into the synthesizer~\cite{alet2021large}.

To synthesize C programs from input-output examples only, we propose \ours{}, which generates the program in a recurrent and token-by-token manner. As the first contribution on model architectures for program synthesis, we propose to use two latent \emph{parallel representations} in the recurrent model. One representation is learned from regular recurrent models as in autoregressive language models~\cite{hochreiter1997long}, with the double attention mechanism over IO pairs proposed in RobustFill~\cite{devlin2017robustfill} and an operation predictor that models the arithmetic relationship between the program input and output. The second representation, named \emph{Latent Execution Trace (\laet{})}, models the hypothetical input signal for the remaining partial program to execute to get to the desired output. Motivated by the line of work on execution-guided program synthesis~\cite{sun2018neural,Ellis2019WriteEAExtendExecution,Zohar2018AutomaticPSExtendExecution,chen2018execution}, we learn a latent representation for C programs which are not executed via interpreters, and train the model given only IO pairs without the intermediate program execution states. The two parallel representations are trained end-to-end. 

\iffalse
To extract potentially complicated relationship between input and outputs, we use an external pre-computed addition and subtraction table with attention mechanism to guide generation of next token. 
\fi

\iffalse
The program is generated in a token-by-token manner: at each time step $t$, the \emph{latent executor} $\phi(\cdot)$ takes the learned representation $h_t$ and the previously generated token $a_t$ as the inputs, and generate $h_{t+1} = \phi(h_t, a_t; IO)$ as the next latent representation of the partial program (here $IO$ is the IO pairs). This enables the algorithm to ``imagine'' how the partial program looks like and how it is related to the output, and thus guides the synthesis effectively. While for program with ground truth interpreter (e.g., Karel), we train $\phi$ to make sure $h_t$ can reconstruct the current execution trace, for C program, we only use the supervision that the last representation $h_T$ should reconstruct the output, and train the entire pipeline in an end-to-end manner. 
\fi

\iffalse
With latent execution trace, \ours{} can generate not only sequential program but also programs with branching and loops. 
\fi

As the second contribution on dataset construction, we demonstrate that it is possible to automatically construct a C codebase that is of high quality, controllable and concise through our proposed program synthesis procedure. Specifically, starting from randomly generated C programs that might contain a lot of redundant statements, we show that via \emph{iterative retraining}, the subsequent generated code from our learned model becomes more concise and similar to human-written ones. Moreover, learning directly from the generated code leads to better performance given the same amount of samples, and improves the sample efficiency. We observe similar results when applying our iterative retraining technique to Karel~\cite{bunel2018leveraging}, another programming by example benchmark consisting of randomly generated programs. Although the initial Karel dataset includes a large proportion of complicated programs with different control flow constructs, we demonstrate that nearly half of the problems can be solved by straight-line programs, which again confirms that randomly generated programs tend to be unnecessarily complicated. We envision that the iterative retraining procedure could greatly reduce laborious efforts in human codebase collection in future research. 

As the third contribution, we show for the first time that short C code in a restricted domain (tens of tokens, no library call) with sequential, branching, loop and simple arithmetic operations can be effectively synthesized from IO pairs only. In particular, while \ours{} tends to generate more concise programs (and does not have exact token match with random generated ground truth code), when measuring whether the program execution outputs match the IO pairs, \ours{} achieves $55.2\%$ accuracy, and outperforms existing neural program synthesis models by around $20\%$. These results demonstrate the effectiveness of learning latent execution traces.

%Second, we are the first to show that a learned latent space plus a planning procedure plays a critical role in program synthesis. Our contribution is three-fold. First, w
\vspace{-0.1in}
\section{Neural Program Synthesis from Input-Output Examples}
\vspace{-0.1in}
In programming by example tasks, the program specification is a set of input-output examples~\cite{devlin2017robustfill,bunel2018leveraging}. Specifically, we provide the synthesizer with a set of $K$ input-output pairs $\{(I^{(k)}, O^{(k)})\}_{k=1}^K$ ($\{IO\}^K$ in short). These input-output pairs are annotated with a ground truth program $P^\star$, so that $P^\star(I^{(k)})=O^{(k)}$ for any $k \in \{1, 2, ..., K\}$. To measure the program correctness, we include another set of held-out test cases $\{IO\}_{test}^{K_{test}}$ that differs from $\{IO\}^K$. The goal of the program synthesizer is to predict a program $P$ from $\{IO\}^K$, so that $P(I)=P^\star(I)=O$ for any $(I, O) \in \{IO\}^K + \{IO\}_{test}^{K_{test}}$.

%\label{sec:c-data}
\textbf{C Program Synthesis}. In this work, we make the first attempt of synthesizing C code in a restricted domain from input-output examples only, and we focus on programs for list processing. List processing tasks have been studied in some prior works on input-output program synthesis, but they synthesize programs in restricted domain-specific languages instead of full-fledged popular programming languages~\cite{balog2016deepcoder,odena2020learning,odena2020bustle}. 

Our C code synthesis problem brings new challenges for programming by example. Compared to domain-specific languages, the syntax and semantics of C are much more complicated, which significantly enlarges the program search space. Meanwhile, learning good representations for partially decoded programs also becomes more difficult. In particular, prior neural program synthesizers that utilize per-line interpreters for the programming language to guide the synthesis and representation learning~\cite{chen2018execution,shin2018improving,nye2020representing,Ellis2019WriteEAExtendExecution,odena2020bustle} are not directly applicable to C. Although it is possible to dump some intermediate variable states during C code execution~\cite{campbell2012executable}, since partial C programs are not executable, we are able to obtain all the execution states only until a full C code is generated, which is too late to include them in the program decoding process. In particular, the intermediate execution state is not available when the partial program is syntactically invalid, and this happens more frequently for C due to its syntax design.
\begin{figure}
    \centering
    \includegraphics[width=\textwidth]{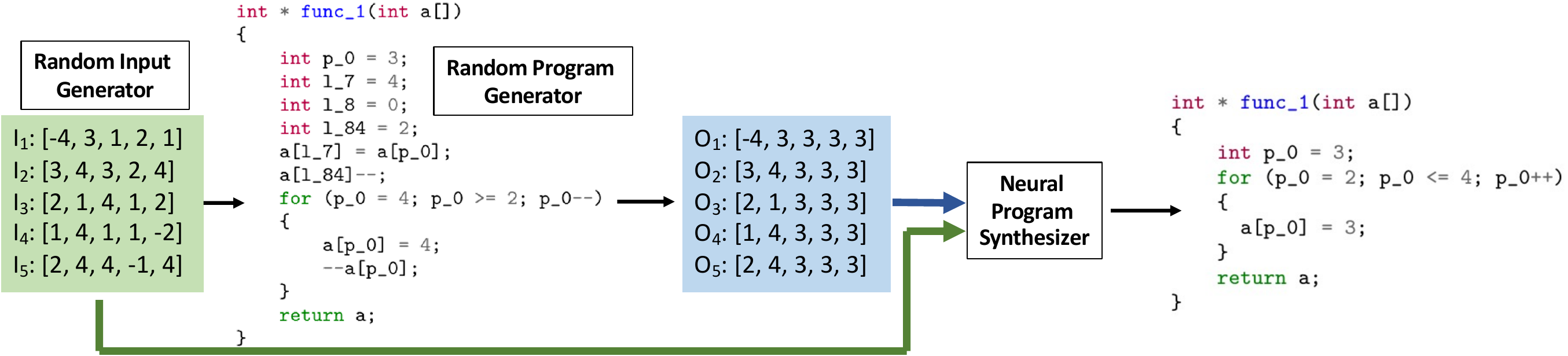}
\caption{\small Illustration of the C program synthesis pipeline. For dataset construction, we develop a random program generator to sample random C programs, then execute the program over randomly generated inputs and obtain the outputs. The input-output pairs are fed into the neural program synthesizer to predict the programs. Note that the synthesized program can be more concise than the original random program.}
\label{fig:ex-c}
\end{figure}

\begin{figure}
    \centering
    \includegraphics[width=\textwidth]{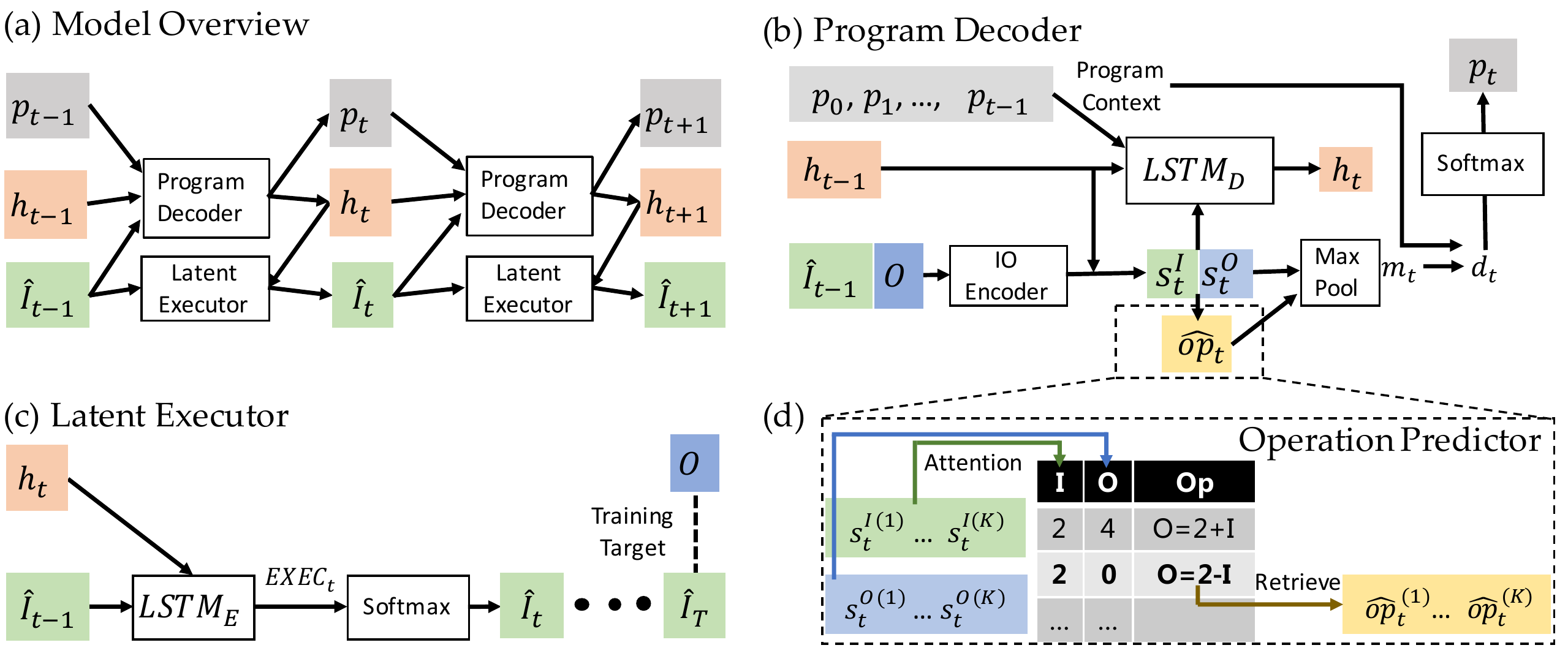}
    \caption{\small (a) An overview of \ours{} model architecture. (b), (c), and (d) present the details of the program decoder, latent executor, and the operation predictor. Note that the operation predictor is specialized for numerical calculation, and thus is not used for the Karel domain.}
    \label{fig:model-architecture}
    \vspace{-1em}
\end{figure}

\def\cL{\mathcal{L}}

\section{Program Synthesis with Learned Execution}
\vspace{-0.1in}
\label{sec:model-architecture}
%\yuandong{Follow the story of introduction and rewrite it. Provide a high-level introduction about different components of the architecture. Emphasize the core design which is neural executor.}
In this section, we present \ours{} which learns to represent the execution of partial programs to guide the synthesis process. Fig.~\ref{fig:model-architecture}(a) provides an overview of \ours{} model architecture which consists of two components, the \emph{program decoder} and the \emph{latent executor}. We present the core design below, and defer more details to Appendix~\ref{app:model-architecture} and Appendix~\ref{app:implementation-details}. 

\subsection{Model Overview}
\vspace{-0.1in}
At a high level, the program decoder (Fig.~\ref{fig:model-architecture}(b)) takes a latent vector $h_{t-1}$ that represents the generated partial program, the previous (generated) program token $p_{t-1}$, and outputs the latent vector $h_t$ and the next program token $p_t$ to be generated at time step $t$: 
\begin{equation}
(h_t, p_t) = \mathrm{ProgramDecoder}(h_{t-1}, p_{t-1}; IO_{t-1}) \label{eq:program-decoder}
\end{equation}
Here the recurrent model is conditioned on the IO pair $IO_{t-1}$. When $IO_t = IO := (I, O)$ for every $t$, i.e., $IO_t$ remains \emph{constant} over the entire recurrent generation process, Eqn.~\ref{eq:program-decoder} represents the standard recurrent architecture used in most autoregressive natural language models~\cite{hochreiter1997long,vaswani2017attention}, and is also used in prior works on program synthesis from input-output examples~\cite{devlin2017robustfill,bunel2018leveraging}. 

For program decoding, the decoder first takes two attention vectors $s_t^I$ and $s_t^O$ computed from IO pairs and latent vector $h_{t-1}$ via double attention~\cite{devlin2017robustfill}, and utilizes a max pooling layer to compute an aggregated embedding $m_t$ for all IO pairs
(Fig.~\ref{fig:model-architecture}(b)):
\begin{equation}
m_t=\mathrm{MaxPool}_{j \in \{1,2,...,K\}}(\mathrm{tanh}(W[s_t^{I(j)}; s_t^{O(j)}])) \label{eq:wo-operation-predictor}
\end{equation}
Here the superscript $(j)$ indicates that the representation is for the $j$-th IO pair, $[a; b]$ is vector concatenation of $a$ and $b$, and $W$ is a trainable matrix. To facilitate the prediction of long programs, we compute an attention vector $d_t$ over previously generated program tokens using the standard attention mechanism~\cite{bahdanau2014neural,luong2015effective}: 
\begin{equation} 
d_t=\mathrm{Attention}(m_t, \{p_0, ..., p_{t-1}\}) \label{eq:d-attention}
\end{equation}
Finally, the next token $p_t$ is sampled from $\mathbb{P}[p_t]=\mathrm{Softmax}(Vd_t)_{p_t}$ where $V$ is a trainable matrix.

\subsection{Latent Executor Design}
\vspace{-0.1in}
As shown in our experiments (Sec.~\ref{sec:exp}), the standard program decoder architecture may not be able to achieve strong performance in program synthesis when the program complexity increases. One main reason is that the standard program decoder only takes the initial IO pairs as the input without considering the program execution, thus the learned representation for the partial program does not effectively guide the synthesis process. Motivated by prior works that utilize execution traces for Karel program synthesis~\cite{chen2018execution,shin2018improving,sun2018neural}, in this paper, we introduce \emph{latent executor} (Fig.~\ref{fig:model-architecture}(c)) which maintains a second representation $\hat I_t$ during program decoding. Intuitively, $\hat I_{t-1}$ models the \emph{hypothetical input} of the partial program $p_{t\ldots T}$ so that its output becomes $O$. Given the estimated input $\hat I_{t-1}$ and the latent vector $h_t$, the latent executor returns $\hat I_{t}$ at the next time step $t$: 
\begin{equation}
    \hat I_t = \mathrm{LatentExecutor}(\hat I_{t-1}, h_t)
\end{equation}
The collection of $\{\hat I_t\}_{t=0}^T$ is the \emph{latent execution trace (\laet{})}. With the help of latent executor, we now use the IO pairs $IO_{t-1} := (\hat I_{t-1}, O)$ instead of $(I, O)$ for the program decoder (Eqn.~\ref{eq:program-decoder}).

%\yuandong{maybe we can mention more detailed observations why they do not work well?} \xinyun{Added.}
\subsection{End-to-end Training}
\vspace{-0.1in}
We train our model with supervised learning, by minimizing the sum of token prediction loss $\cL_{Prog}$, and the latent executor loss $\cL_{Exec}$: 
\begin{equation}
\cL=\cL_{Prog} + \cL_{Exec}
\end{equation}
Specifically, $\cL_{Prog} :=\sum_{t=1}^T\mathrm{Loss}(p_t, p^\star_t)$ is the step-by-step cross-entropy loss between the predicted programs $p_{1\ldots T}$ and the ground truth programs $p^\star_{1\ldots T}$.

For latent executor, since the semantics of partial programs (e.g., partial C programs) are not always well-defined, there is no step-by-step training supervision. However, the output of the executor should be consistent with the program specification after taking the annotated ground truth program as the input. Therefore, we set $\hat I_0 = I$ (true input) and minimize the distance between $\hat I_T$ and $O$ (true output) after the program finishes:
\begin{equation}
\cL_{Exec}=\mathrm{Loss}(\hat{I}_T, O)
\end{equation}
Note that $\cL_{Exec}$ does not rely on any assumptions of the partial program semantics, and thus is applicable to both domain-specific languages and general-purpose programming languages such as C. In our evaluation, equipping with the latent executor significantly improves the program prediction performance, where each program could include up to 256 tokens. \yuandong{People may be curious about whether this training would break if the program is too long? What are typical length of a program in the training set.} \xinyun{Added. The length of random programs is much larger, but the decoded programs mostly have around 50 tokens.} \yuandong{In the C section, the maximal token length is 256. So what happens to if the program is much longer than that?}

\def\cD{\mathcal{D}}

\subsection{Data Regeneration and Iterative Retraining}
\label{sec:iterative-retraining}
Interestingly, once our model is trained on the initial random generated programs $\cD_0$, the predicted program becomes more concise and resembles human-written code. While the exact token match accuracy is low even on the training set, the model still satisfies the IO pairs for many problems. We leverage such a phenomenon to construct a new dataset $\cD_1$ with higher-quality programs from $\cD_0$. Specifically, we run beam search on the trained model to predict program $p_{0\ldots T}$ given input-output pairs in the training set. If model prediction $p_{0\ldots T}$ satisfies all the input-output examples and held-out cases, we replace the original program $p^\star_{0\ldots T}$ with $p_{0\ldots T}$ in $\cD_1$, and keep $p^\star_{0\ldots T}$ otherwise. Afterward, we re-train the model on $\cD_1$. In Sec.~\ref{sec:exp}, we will demonstrate that the retraining process further improves the model performance, especially with smaller training datasets.

\section{Restricted C Program Synthesis Domain}
\vspace{-0.1in}

\begin{table}[t]
\centering
\caption{\small The comparison between our restricted C domain and existing programming by example tasks.}
\label{tab:task-comparison}
\begin{tabular}{ccccc}
\toprule
& Control flow  & Variables & Arithmetics & No helper functions \\
\midrule
Restricted C (Ours) & \cmark & \cmark & \cmark & \cmark \\
Karel~\cite{bunel2018leveraging} & \cmark & $-$ & $-$  &  $-$  \\
DeepCoder~\cite{balog2016deepcoder} &  $-$  & \cmark & \cmark &  $-$  \\
FlashFill~\cite{gulwani2011automating} &  $-$  &  $-$  &  $-$  &  $-$  \\
\bottomrule
\end{tabular}
\end{table}

In this section, we discuss our restricted C program synthesis domain, and our operation predictor design for improving the numerical reasoning ability of program synthesis models.

\subsection{Data Generation}
\vspace{-0.1in}

Collecting large-scale high-quality datasets for program synthesis requires a lot of human efforts, and we aim to reduce the manual work for dataset construction.

Our data generator is built upon Csmith~\cite{yang2011finding}, a random C code generation tool originally designed for finding bugs in compilers. Following the common practice of generating input-output pairs, for each program, we randomly sample 5 numerical lists as the program inputs, and execute the program to obtain the corresponding output lists. This is similar to existing works on PBE problems that sample programs based on a probabilistic context-free grammar, randomly generate valid inputs for the programs and obtain the outputs~\cite{parisotto2016neuro,devlin2017neural,balog2016deepcoder}. This creates infinite samples for synthesizing programs in domain-specific languages. While the programs sampled in this way differ from human-written code, Sec.~\ref{sec:iterative-retraining} shows that they can be converted to be more concise and human-like.

\textbf{The subset of language features used}. Our generated program has variable declaration, variable assignment, and expressions with addition or subtraction operations. The programs also have non-sequential statements, including \texttt{If} statements, \texttt{For} loops, \texttt{Continue} and \texttt{Break} statements. Except for the input argument which is a list, all variables declared are integers, and all program statements are integer manipulation. Each expression has at most 2 mathematical operations, and chaining the full C program could perform multi-step numerical calculation (e.g., \texttt{p0 = p0 - p1 + p2; p0 = p0 - 1;}). Looping statements other than \texttt{For} (i.e., \texttt{While} or \texttt{Do-While} loops) are not supported. Note that we only constrain the final program length ($\le 256$ tokens) and the program can have nested for-loops and complicated if-conditions. 

\textbf{Post-processing}. We perform a few post-processing steps to obtain our final programs from programs generated by Csmith (see Fig.~\ref{fig:ex-c} for an example). We resample different components of the program, so that (1) each constant numerical value lies in $[-4, 4]$, (2) mathematical operators only contain addition and subtraction, and (3) upper/lower limits of \texttt{For} loops are positive and within the length of the list. Programs are discarded if they are trivial (e.g., constant or identity mappings), or the input-output examples include values out of the range $[-4, 4]$. \yuandong{Reviewer may complain that $[-4,4]$ is too restrictive. So we might want to give like 2 pages of generated examples in supplementary materials?} 

\textbf{Final dataset}. We reweight the program distribution so that at least half of them include \texttt{For} loops. Our full dataset includes $500K$ samples in the training set, $1K$ samples in the validation set, and $1K$ samples in the test set. As shown in Fig.~\ref{fig:ex-c}, the randomly sampled program may contain redundant statements, which can be easily avoided by human programmers. We compare our restricted C domain to prior datasets of programming by example in Table~\ref{tab:task-comparison}.

\subsection{Program Decoding with the Operation Predictor}
\vspace{-0.1in}
For program decoder, predicting the next program token $p_t$ is non-trivial, especially when mathematical reasoning is required~\cite{saxton2019analysing,lample2019deep}. To improve the program synthesis performance for domains involving numerical calculation, such as our restricted C domain, we design an associative memory structure named \emph{operation predictor} (Fig.~\ref{fig:model-architecture}(d)), based on the following intuition: given the input $I=2$ and output $O=4$, human would infer that ``$O=I+2$'' might be the desired operation and write down the code accordingly. To materialize such an intuition, we create a pre-computed table that covers all possible integer addition and subtraction operations for valid input and output list values. We defer the details of the model architecture to Appendix~\ref{app:operation-predictor}. The program decoding process remains similar to the one described in Sec.~\ref{sec:model-architecture}, and we highlight the key differences as follows. 

The operation predictor takes two attention vectors $s_t^I$ and $s_t^O$ as the representations of input-output examples, and yields an operator embedding $\hat{op}_t$. To compute the aggregated embedding vector for all input-output examples, we modify Eqn.~\ref{eq:wo-operation-predictor} to also take $\hat{op}_t$ as an input of the max pooling layer:
\begin{equation}
m_t=\mathrm{MaxPool}_{j \in \{1,2,...,K\}}(\mathrm{tanh}(W[s_t^{I(j)}; s_t^{O(j)}; \hat{op}_t^{(j)}])) \label{eq:with-operation-predictor}
\end{equation}

To train the operation predictor, we add an additional loss $\cL_{Op}$: 
\begin{equation}
\cL=\cL_{Prog} + \cL_{Exec} + \cL_{Op}
\end{equation}
$\cL_{Op}$ is designed to ensure that the operation predictor predicts operations related to IO pairs, and we defer the details to Appendix~\ref{app:operation-predictor}.

\textbf{Limitations.} In our current implementation of the operation predictor, the operation table is only able to enumerate the arithmetic operations over a pre-defined constant set, thus it requires that the set of possible numerical values in input-output pairs is finite. One way of extending our operation predictor to support potentially unbounded numerical calculation is to combine it with the subword tokenizer, which has been commonly used in recent language models~\cite{devlin2018bert,chen2021evaluating,austin2021program}. We consider designing general-purpose number representation for better mathematical reasoning as future work.

\section{Experiments}
\vspace{-0.1in}
\label{sec:exp}
In this section, we discuss our results on synthesizing programs in Karel and C languages. We first show that {\ours} achieves competitive performance on Karel benchmark. Then we present the results on our restricted C benchmark, and demonstrate that our approach significantly outperforms existing neural program synthesis models. Finally, we discuss the effect of iterative retraining.

\begin{table}[t]
\caption{\small The comparison between {\ours} and baseline neural program synthesis models in our evaluation.}
\label{tab:approach-comparison}
\scalebox{0.9}{
\begin{tabular}{lc|ccc|cc}
\toprule
& \multirow{2}{*}{\ours}  & Exec & Shin et al. & Bunel et al. & RobustFill & Property Signatures \\
& & \cite{chen2018execution} & \cite{shin2018improving} & \cite{bunel2018leveraging} & \cite{devlin2017robustfill} & \cite{odena2020learning}  \\
\midrule
$+$ Program execution & \cmark & \cmark & \cmark &  $-$  &  $-$  &  $-$   \\
No interpreter needed & \cmark  &  $-$  &  $-$  & \cmark & \cmark & \cmark  \\
\bottomrule
\end{tabular}}
\end{table}

\vspace{-0.1in}
\subsection{Karel Program Synthesis}
\subsubsection{Evaluation Setup}
\label{sec:exp-setup-karel}

\textbf{Karel domain.} Karel is an educational programming language~\cite{pattis1981karel}, and has been studied in recent works on neural program synthesis from input-output examples~\cite{devlin2017neural,bunel2018leveraging,chen2018execution,shin2018improving}. A Karel program controls a robot in a 2D grid world. There are instructions that control the robot, e.g., \texttt{move}, \texttt{turnLeft} and \texttt{PutMarker}, as well as conditionals and loops, i.e., \texttt{if}, \texttt{repeat} and \texttt{while}. See Appendix~\ref{app:karel-details} for grammar specification and the state representation. 

We train and evaluate all models on the Karel dataset introduced in~\cite{bunel2018leveraging}. The dataset contains randomly sampled programs from the Karel DSL ($1.1M$ training samples, $2.5K$ samples in the validation set and $2.5K$ samples in the test set). Each program includes 5 input-output pairs as the specification, and the sixth pair as the held-out test case. Following the prior work, we evaluate two metrics: (1) {\bf Exact Match}: the predicted program is the same as the ground truth; (2) {\bf Generalization}: the predicted program satisfies both the input-output pairs and the held-out input-output test case.

\textbf{Baselines.} \emph{Bunel et al.}~\cite{bunel2018leveraging} designed the first program synthesis model for the Karel benchmark with a similar high-level design as RobustFill, but they use convolutional neural networks (CNN) to encode the Karel grid maps. Compared to {\ours}, this model does not utilize any program execution information, and does not include our latent executor. Instead of directly synthesizing the program from input-output examples, the model in \emph{Shin et al.}~\cite{shin2018improving} first predicts the execution traces containing the robot actions from the input-output pairs, then decodes the program based on the execution traces. This model improves the prediction performance over Bunel et al., but it requires the full execution traces for model training and an interpreter for execution. \emph{Exec}~\cite{chen2018execution} leverages the execution states of partial generated programs to guide the subsequent synthesis process, but the execution states are obtained from the Karel interpreter rather than learned by the model, thus this approach represents the ideal scenario where the partial programs could be executable.

Our model architecture for Karel is largely similar to the model for C code synthesis, except that we employ the CNN encoder in Bunel et al.~\cite{bunel2018leveraging} in our program decoder and latent executor. The comparison with baseline models is shown in the middle block of Table~\ref{tab:approach-comparison}. All models use the beam search for decoding programs, with the beam size of 64.
%Specifically, we use 2 convolutional neural networks to encode the input and output grids respectively, then apply another convolutional block to produce the vectors for the program decoder and the latent executor. 

\vspace{-0.1in}
\subsubsection{Results}
\label{sec:karel-results}

\begin{table}[t]
\begin{minipage}{0.45\linewidth}
\caption{\small Results on Karel dataset. \textbf{Gen} and \textbf{Exact} denote generalization and exact match accuracies.}
\label{tab:karel}
\begin{tabular}{lcc}
\toprule
\textbf{Approach}  & \textbf{Gen} & \textbf{Exact} \\
\midrule
\ours{} & 83.68\% & 41.12\% \\
Exec~\cite{chen2018execution} & \textbf{86.04\%} & 39.40\% \\
Bunel et al.~\cite{bunel2018leveraging} & 77.12\% & 32.17\% \\
Shin et al.~\cite{shin2018improving} & 81.30\% & \textbf{42.80\%} \\
\bottomrule
\end{tabular}
\end{minipage}
\begin{minipage}{0.5\linewidth}
    \centering
    \includegraphics[width=\linewidth]{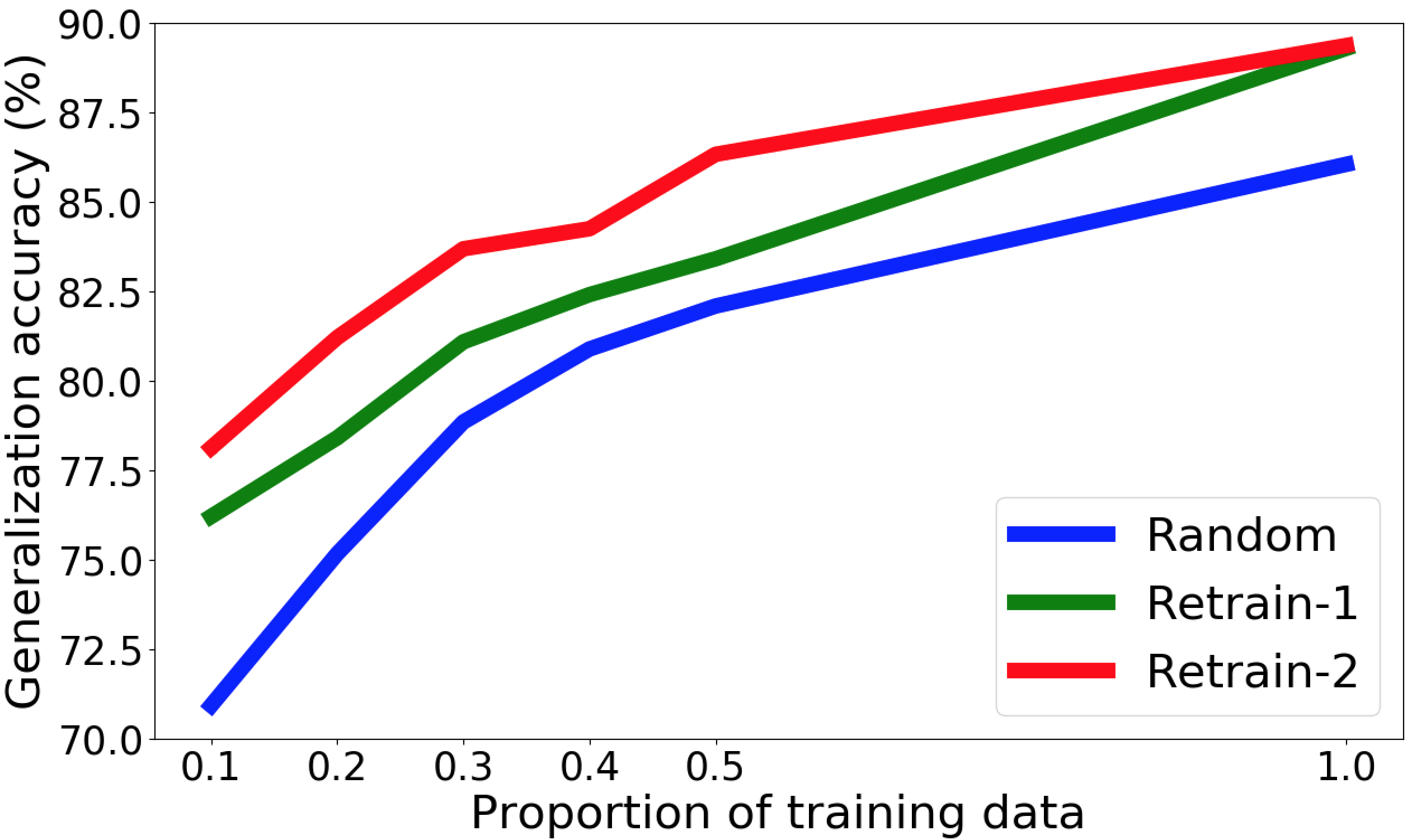}
    \captionof{figure}{\small Generalization accuracies with different training data sizes on Karel. With the full training set, the accuracies are $86.04\%$, $89.28\%$ and $89.36\%$ for training on random programs, retraining for 1 and 2 iterations. \yuandong{Make the line thicker.} \xinyun{Updated.} }
    \label{fig:retrain-karel-acc}
\end{minipage}
\vspace{-0.3in}
\end{table}

\begin{figure}[t]
    \centering
    \begin{subfigure}[t]{0.45\linewidth}
    \includegraphics[width=\linewidth]{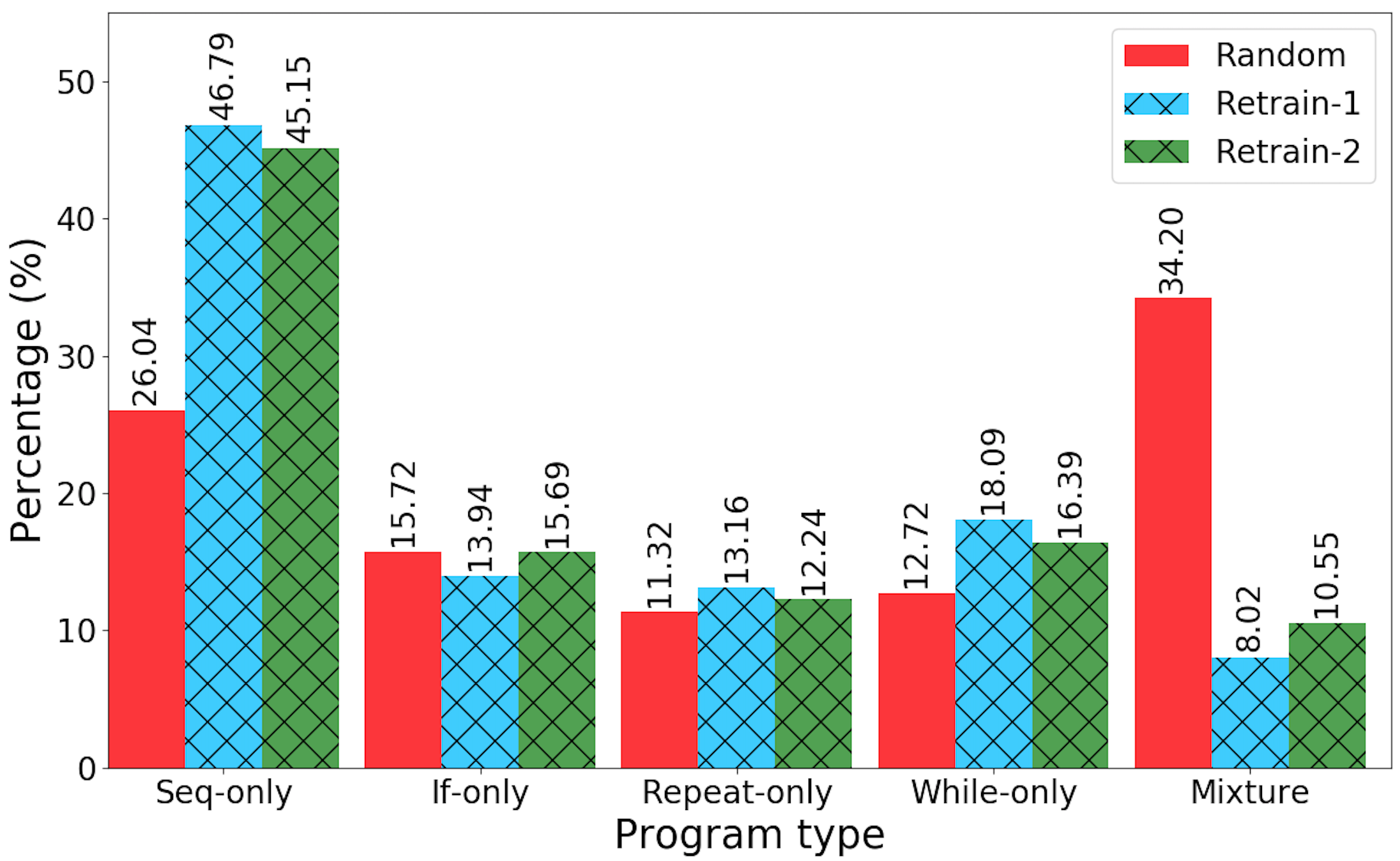}
    \caption{}
    \label{fig:retrain-karel-dist}
    \end{subfigure}
    \begin{subfigure}[t]{0.45\linewidth}
    \includegraphics[width=\linewidth]{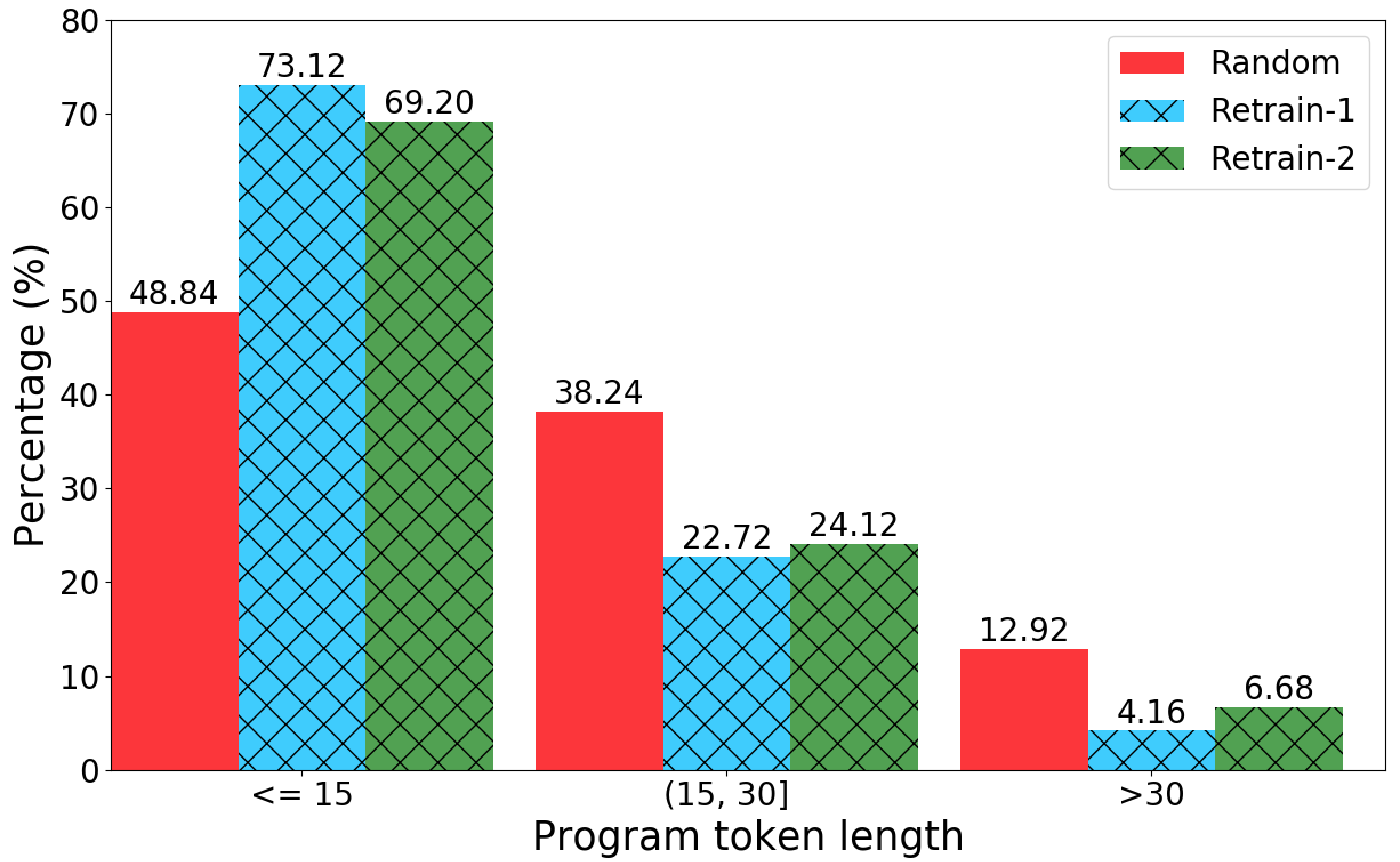}
    \caption{}
    \label{fig:karel-dist-len}
    \end{subfigure}
    \vspace{-0.1in}
    \caption{\small Program distributions after iterative retraining on Karel. (a) The distributions of different program types. \emph{Seq-only}: no control flows. \emph{If-only}: the program includes If statements but no loops. \emph{Repeat/While-only}: the program includes Repeat/While loops, but no other control flow constructs. \emph{Mixture}: the program includes at least two types of control flow constructs. (b) The distributions of programs with different token lengths. \yuandong{What's the difference between xxx-only and mixture?} \xinyun{Explained.}}
    \label{fig:retrain-karel}
    \vspace{-0.2in}
\end{figure}

We present the results of {\ours} and baseline model architectures in Table~\ref{tab:karel}. First, {\ours} outperforms all baselines that do not incorporate the partial program execution information, and achieves competitive performance compared with the Exec algorithm that requires an interpreter to obtain the partial program execution states. In particular, {\ours} achieves a higher generalization accuracy than Shin et al. with lower exact match accuracy, showing that decoded programs by \ours{} are more different from randomly generated programs. Although Shin et al. also model the program execution by predicting the robot actions, the prediction of the action traces does not take the program structure into account, resulting in the inferior performance.
\vspace{-0.1em}
\subsection{C Code Synthesis}
\subsubsection{Evaluation Setup}
\label{sec:exp-c-setup}
Given the variety of C programs, we observe that the exact match accuracies of models are mostly nearly 0. Therefore, we focus on evaluating the generalization accuracy, and we consider the predicted program to be correct when it satisfies both the 5 input-output examples and 5 held-out test cases.

\textbf{Baselines.} We compare the full {\ours} with its multiple ablated versions:
\begin{itemize}[noitemsep,topsep=0pt,parsep=0pt,partopsep=0pt]%,leftmargin=*]
    \item \texttt{NoExecutor}. The program decoder (Eqn.~\ref{eq:program-decoder}) always takes the initial input-output pairs as the input; i.e,. $\hat I_t = I_0$ for every $t$.
    \item \texttt{NoPartialExecutor}. $\hat I_t = I_0=I$ for every $t$ and additionally $h_T$ is regularized so that $\mathrm{LatentExecutor}(I_0, h_T)$ matches the output $O$ under loss $\cL_{Exec}$. Therefore, no partial latent execution. 
    \item \texttt{NoOpPredictor}. The max pooling layer only takes the vectors computed by the double attention as the input (Eqn.~\ref{eq:wo-operation-predictor}).
    \item \texttt{NoAttentionInDecoding}. There is no attention over decoded program tokens, and the output of the max pooling layer is directly fed into the output softmax layer; i.e., $\mathbb{P}[p_t]=\mathrm{Softmax}(Vm_t)_{p_t}$ (compared to Eqn.~\ref{eq:d-attention}).
\end{itemize}
We also compare with existing neural program synthesis models with good performance on related tasks, as shown in the rightmost block of Table~\ref{tab:approach-comparison}. \emph{RobustFill}~\cite{devlin2017robustfill} is the state-of-the-art neural network architecture on FlashFill benchmark, which synthesizes string manipulation programs in a domain-specific language. As described in Sec.~\ref{sec:model-architecture}, the input-output encoder and the program decoder architectures in RobustFill are similar to {\ours}, except that it does not include the latent executor, operation predictor, and the attention on the decoded program sequence. \yuandong{Why we don't use Bunel et al and Shin et al for C program? We want to explain a bit. If the conceptual tables of different methods can be shown, the reasoning would be easier.} \xinyun{I added some discussion in Karel baseline description, do we need more explanation here?}

\emph{Property Signatures}~\cite{odena2020learning} was designed for synthesizing list manipulation programs in domain-specific languages, but instead of taking the raw input and output lists as the neural network input, they design some properties that distinguish different programs, then take the values of these properties as the model input. A sample property could be whether the program output is the same as the input, and the property values could be ``All True'', ``All False'', or ``Mixed'', indicating that the property always holds for any input-output pair in the specification, never holds, and holds for some pairs but not others, respectively. We customize the original design~\cite{odena2020learning} for our setting. First, our property set takes the format of $O = C + I?$ and $O = C - I?$, where $C \in [-4, 4]$. For example, $O = 2 + I?$ means whether the output $O$ could be calculated by adding $2$ to the input $I$. These properties focus more on numerical calculation, similar to our operation predictor. Second, different from the task in~\cite{odena2020learning}, our C programs sometimes manipulate only a subset of the input lists, thus encoding the list with a single property value is inappropriate. Instead, we compute the property value per element in input-output pairs, use a bi-directional LSTM to encode the property values as a sequence, then take the outputs of the bi-LSTM for program prediction.

\subsubsection{Results}

\begin{figure}
\begin{minipage}{0.25\textwidth}
\begin{minted}[fontsize=\scriptsize]{c}
int * func_1(int a[])
{
    int p_0 = 0;
    int l_25 = 4;
    a[p_0] = 1;
    --a[l_25];
    return a;
}
\end{minted}
\end{minipage}
\begin{minipage}{0.33\textwidth}
\begin{minted}[fontsize=\scriptsize]{c}
int * func_1(int a[])
{
    int p_0 = 2;
    int l_12 = 3;
    for (p_0 = 1; p_0 <= 2; p_0++)
    {
        a[p_0]--;
    }
    a[l_12] = a[l_12] + 4;
    return a;
}
\end{minted}
\end{minipage}
\begin{minipage}{0.2\textwidth}
\begin{minted}[fontsize=\scriptsize]{c}
int * func_1(int a[])
{
    int p_0 = 0;
    int l_7 = 3;
    int l_8 = 1;
    a[l_8] = (a[l_7] - a[p_0]);
    for (p_0 = 3; p_0 <= 4; p_0++)
    {
        for (int p_1 = 1; p_1 <= 2; p_1++)
        {
            a[p_1] = a[p_1] + a[p_0];
            a[p_1] = a[p_1] + 2;
        }
    }
    return a;
}
\end{minted}
\end{minipage}
\caption{\small Sample programs that could be correctly predicted by {\ours}, but wrongly predicted by models without the latent executor. These programs require multiple different operations for different input list elements.\yuandong{Any explanation why?} \xinyun{Added.}}
\label{fig:ex-c-exec}
\vspace{-0.15in}
\end{figure}

\begin{table}[t]
\begin{minipage}{0.5\linewidth}
\caption{\small Results on C dataset.}
\label{tab:c}
\centering
\begin{tabular}{lc}
\toprule
\textbf{Approach}  & Accuracy \\
\midrule
\ours{} & \textbf{55.2\%} \\
\texttt{NoAttentionInDecoding} & 53.5\% \\
\texttt{NoOpPredictor} & 53.7\% \\
\texttt{NoPartialExecutor} &  42.9\% \\
\texttt{NoExecutor} &  38.6\% \\
RobustFill~\cite{devlin2017robustfill} & 37.6\% \\
Property Signatures~\cite{odena2020learning} & 34.5\% \\
\bottomrule
\end{tabular}
\end{minipage}
\begin{minipage}{0.45\linewidth}
    \includegraphics[width=\linewidth]{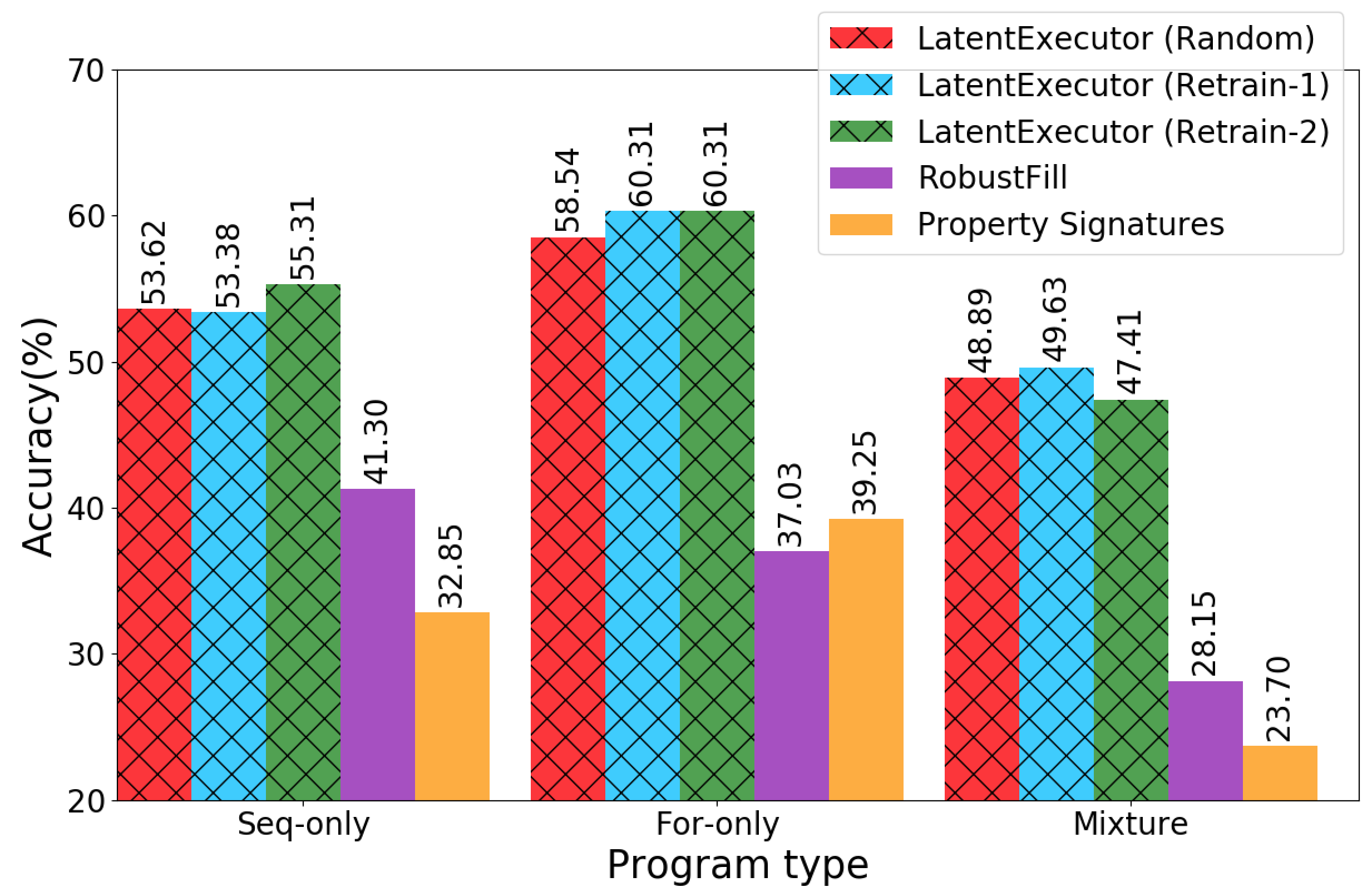}
    \captionof{figure}{\small Accuracies of different program types on C dataset.}
    \label{fig:c-acc-dist}
\end{minipage}
\end{table}
    
\begin{figure}[t]
    \centering
    \vspace{-0.1in}
    \begin{subfigure}[t]{0.45\linewidth}
    \includegraphics[width=\linewidth]{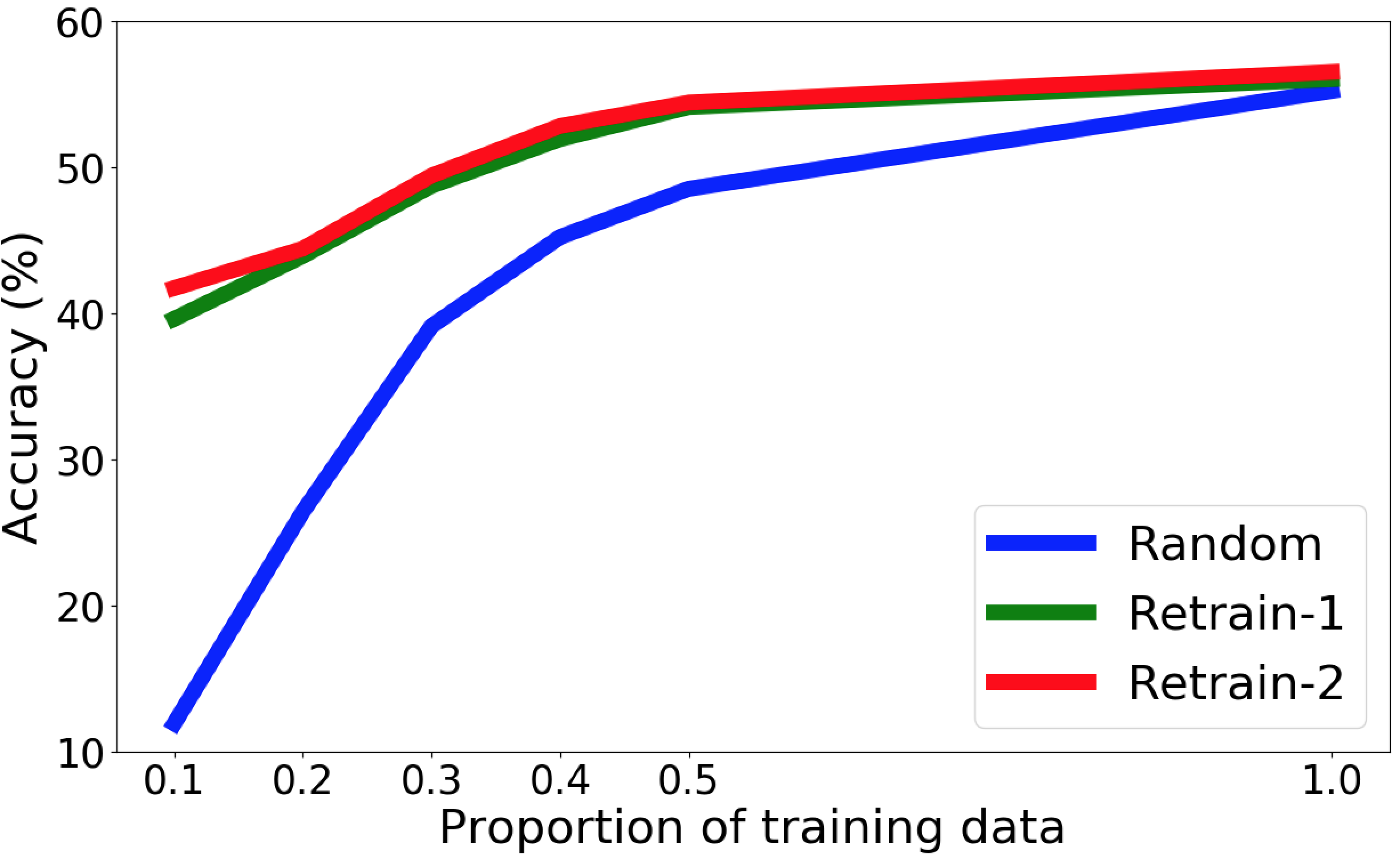}
    \caption{}
    \label{fig:retrain-c-acc}
    \end{subfigure}
    \begin{subfigure}[t]{0.45\linewidth}
    \includegraphics[width=\linewidth]{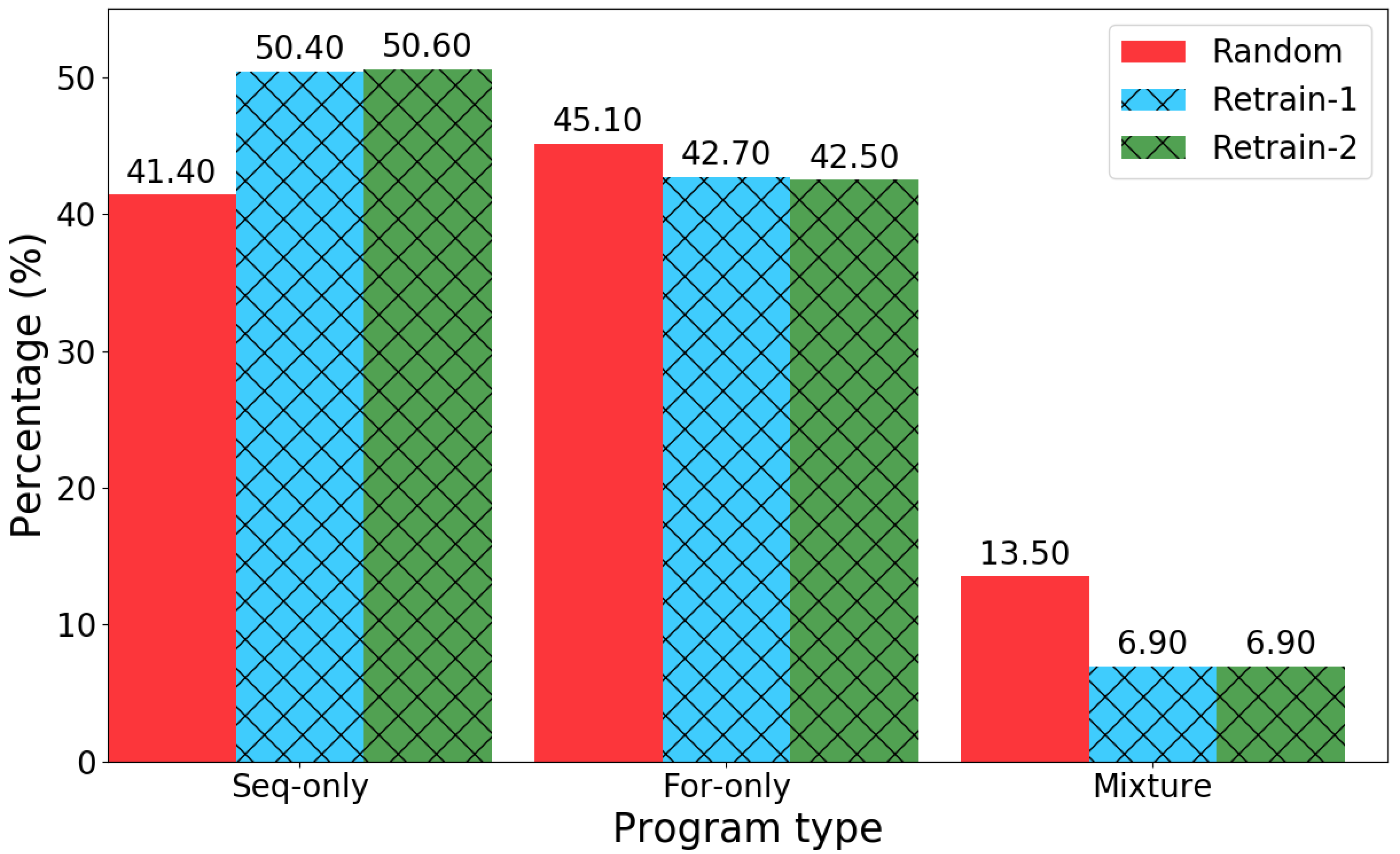}
    \caption{}
    \label{fig:retrain-c-dist}
    \end{subfigure}
    \vspace{-0.1in}
    \caption{\small Results of iterative retraining on the C dataset. (a) Accuracies with different training data sizes. With the full training set, the accuracies are $55.2\%$, $56.0\%$ and $56.5\%$ for training on random programs, retraining for 1 and 2 iterations, respectively. (b) The program distributions after each retraining iteration.}
    \label{fig:retrain-c}
    \vspace{-0.1in}
\end{figure}

\begin{figure}[t]
    \centering
    \begin{subfigure}[t]{0.45\linewidth}
    \includegraphics[width=\linewidth]{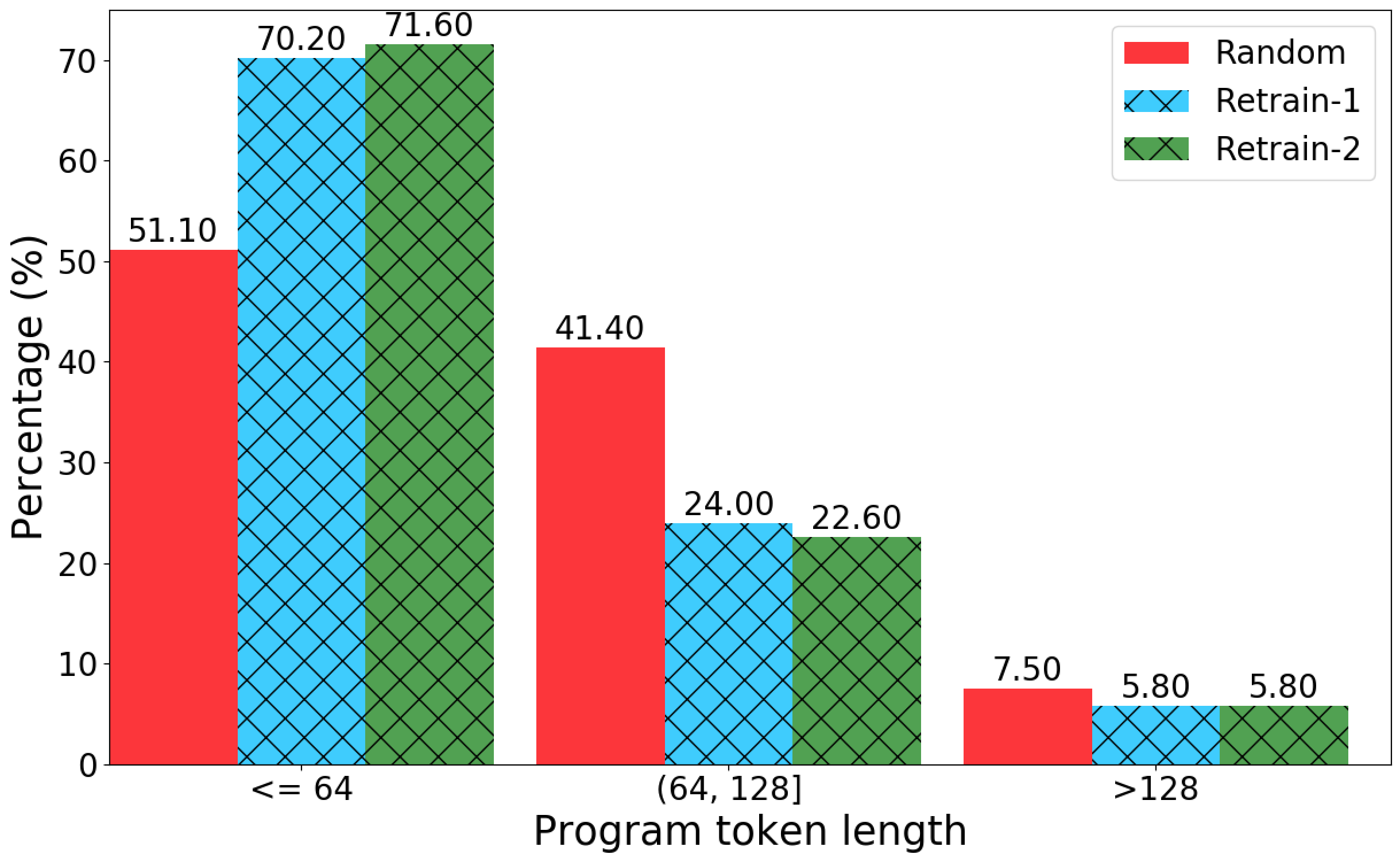}
    \caption{}
    \label{fig:c-dist-len}
    \end{subfigure}
    \begin{subfigure}[t]{0.45\linewidth}
    \includegraphics[width=\linewidth]{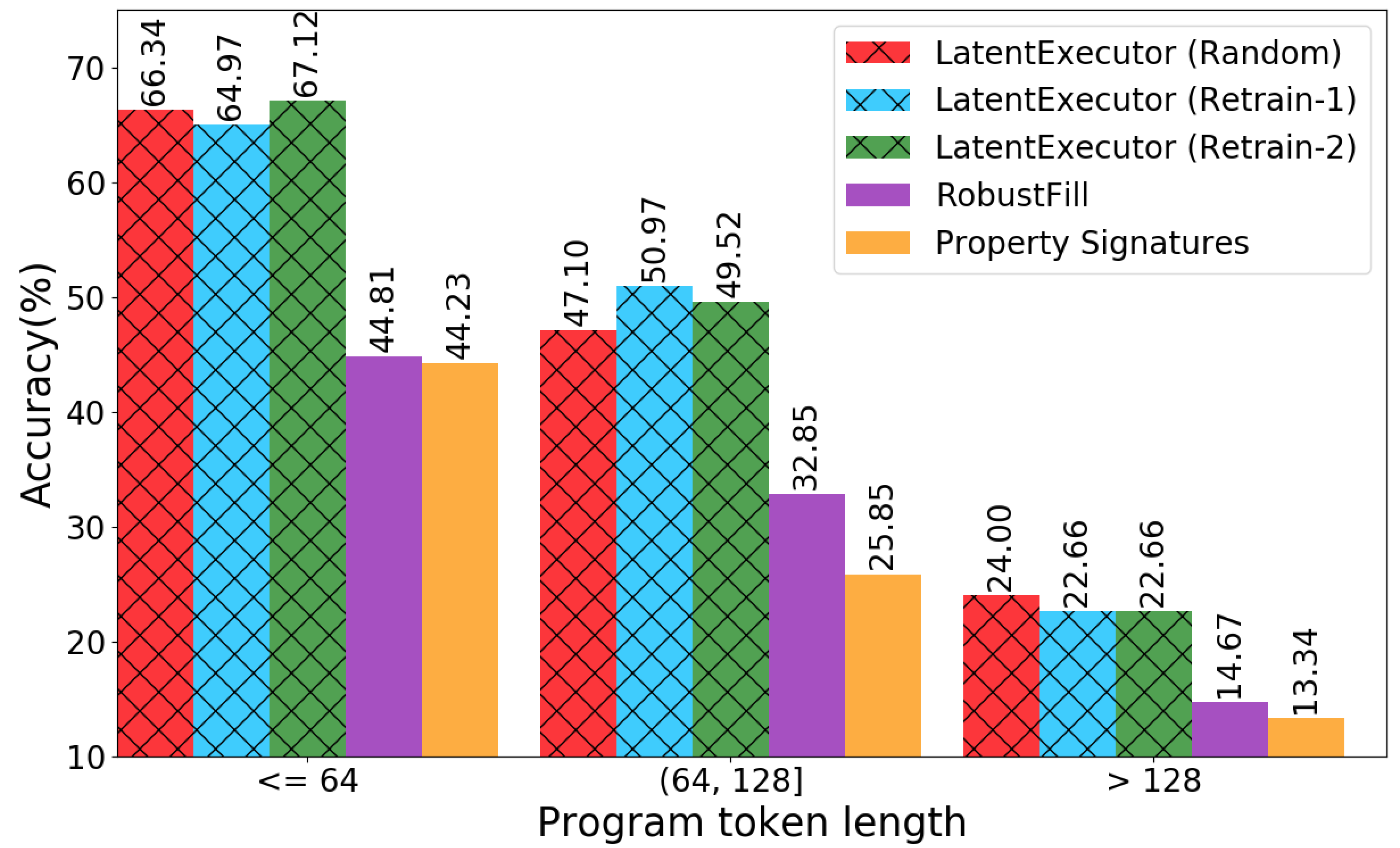}
    \caption{}
    \label{fig:c-len-acc}
    \end{subfigure}
    \vspace{-0.1in}
    \caption{\small Results on programs of different token lengths on the C dataset. (a) The program token length distributions after each retraining iteration. (b) The accuracies on programs of different token lengths.}
    \vspace{-0.2in}
    \label{fig:res-c-len}
\end{figure}

Table~\ref{tab:c} presents the results, where all models are trained on the initial random programs. The full {\ours} outperforms other variants, and improves the performance of RobustFill and Property Signatures by around $20\%$. We also increase the model size of RobustFill to see if the improvement comes from larger model size, but the results are not better. In particular, the latent executor significantly increases the prediction accuracy, and achieves better results than \texttt{NoPartialExecutor}, which shows that learning latent execution traces leads to better partial program representations. In Fig.~\ref{fig:ex-c-exec}, we present sample programs that could be correctly synthesized by {\ours}, but models without the latent executor provide the wrong prediction. We observe that the latent executor is beneficial when the program involves different manipulations for different list elements, e.g., more than one For loop and different mathematical calculations. Our breakdown results on programs of different complexity also justify this observation. We first present the results on programs with different control flow constructs in Fig.~\ref{fig:c-acc-dist}. Specifically, \emph{Seq-only} includes programs with no control flow constructs, \emph{For-only} includes programs with For loops but no If statements, and \emph{Mixture} includes programs with both For loops and If statements. Then we demonstrate the results on different program lengths in Fig.~\ref{fig:c-len-acc}. We show that {\ours} achieves decent performance on long and complicated programs, while the accuracies of baseline models drop dramatically.  \yuandong{One concrete program where Neural Executor works but ``no Neural Executor'' doesn't work would be very interesting to the audience.} \xinyun{Added some examples in Fig.~\ref{fig:ex-c-exec}.}

\vspace{-0.1em}
\subsection{Discussion of Iterative Retraining}
In Fig.~\ref{fig:retrain-karel-acc}, we show the effectiveness of retraining on decoded Karel programs (Sec.~\ref{sec:iterative-retraining}). We observe that retraining for one iteration is sufficient, and it significantly improves the generalization accuracy by over $3\%$. To understand the differences between predicted programs and randomly generated programs, we demonstrate the changes of dataset distributions after each retraining iteration in Fig.~\ref{fig:retrain-karel-dist} and~\ref{fig:karel-dist-len}. We observe that the model learns to predict more concise programs than the ground truth for a large proportion of input-output examples, and considerably alters the dataset distribution so that it becomes more concentrated on short programs with simplified control flow structures. Specifically, from Fig.~\ref{fig:retrain-karel-dist}, although the initial Karel dataset seems to include a large proportion of complicated programs with different control flow constructs, our model synthesizes straight-line programs for nearly half of the samples, which means that many loops and branches in the annotated ground truth programs are unnecessary. This distribution shift also explains the gap between the exact match and generalization accuracies. The program distribution after the second retraining iteration is largely similar to the first iteration, thus retraining for more iterations does not considerably improve the performance. Note that in the second iteration, the synthesizer tends to generate slightly more complicated programs than the first iteration, in order to deal with the cases when the input-output examples oversimplify the intended program functionality. For example, sometimes the input-output examples do not cover the edge cases that the robot may encounter, thus adding additional \texttt{If} branches could avoid the crashes when testing on held-out cases.

Fig.~\ref{fig:retrain-c-acc} presents the results of retraining on decoded C programs. Similarly, retraining improves the prediction accuracy, especially when the training set is small. From Fig.~\ref{fig:retrain-c-dist} and~\ref{fig:c-dist-len}, we again observe that the model tends to predict shorter programs than the random code, and it eliminates unnecessary control flows to simplify the programs. We present more examples in Appendix~\ref{app:retrain}.

\section{Related Work}
\label{sec:work}

\textbf{Programming by example.} Programming by example problems have been widely studied with various applications, and recent works have developed deep neural networks as program synthesizers~\cite{gulwani2012spreadsheet,parisotto2016neuro,devlin2017robustfill,bunel2018leveraging}. Most prior works focus on synthesizing programs in domain-specific languages, such as FlashFill~\cite{parisotto2016neuro,devlin2017robustfill,vijayakumar2018neural} for string transformation, Karel~\cite{bunel2018leveraging,shin2018improving,chen2018execution,gupta2020synthesize} for simulated robot navigation, and LISP-style languages for list manipulation~\cite{balog2016deepcoder,polosukhin2018neural,Zohar2018AutomaticPSExtendExecution,nye2019learning}. In this work, we make the first attempt of synthesizing C code in a restricted domain from input-output examples only, and we focus on the list manipulation domain.

Some recent works investigate the limitations of synthetic datasets and the ambiguity in program specifications for neural program synthesis~\cite{shin2019synthetic,clymo2020data,suh2020creating,laich2019guiding}. These works focus on reducing the bias of data distributions and generating more diverse input-output pairs, while our data regeneration aims to improve the quality of programs. We consider incorporating both lines of work to further improve the dataset quality as future work. In addition, drawing the inspiration from self-training and bootstrapping techniques developed for other applications~\cite{mooney1993bootstrapping,abney2002bootstrapping,mcclosky2006effective,xie2020self} to extend our iterative retraining scheme is also another future direction.

\textbf{Execution-guided program synthesis.} To learn better program representations, some recent works incorporate the execution information to guide the synthesis process~\cite{sun2018neural,Zohar2018AutomaticPSExtendExecution,shin2018improving,chen2018execution,Ellis2019WriteEAExtendExecution,tian2019learning,balog2020neural,gupta2020synthesize,odena2020bustle,nye2020representing,mandal2021learning}. In particular, leveraging partial program execution states improves the performance for several program synthesis tasks~\cite{chen2018execution,Zohar2018AutomaticPSExtendExecution,Ellis2019WriteEAExtendExecution,nye2020representing}. However, existing approaches rely on program interpreters to provide the intermediate execution results whenever applicable. In contrast, we demonstrate that our latent executor learns the latent execution traces (\laet{}) without such a requirement. Besides program synthesis, execution traces have also been utilized for other software engineering applications~\cite{alam2019zero,mendis2019ithemal}.

\textbf{Neural execution.} Our latent executor is related to prior works on learning to execute algorithms~\cite{zaremba2014learning,velivckovic2019neural,yan2020neural} and programs~\cite{bieber2020learning}. They focus on predicting execution results for full algorithms and programs, but do not utilize them for program synthesis. Latent state prediction has also been studied in other applications such as task-oriented dialogue systems~\cite{min2020dsi,zhang2020probabilistic} and robotics~\cite{paxton2019prospection}.

\section{Conclusion}
\label{sec:conc}
\vspace{-0.1in}
In this work, we propose \ours{}, which learns the latent representation to approximate the execution of partial programs, even if their semantics are not well-defined. We demonstrate the possibility of synthesizing elementary C code from input-output examples only, and leveraging learned execution significantly improves the prediction performance by around $20\%$. Meanwhile, compared to the randomly generated programs, \ours{} synthesizes more concise programs that resemble human-written code, and training on these synthesized programs further improves the prediction performance for both Karel and C program synthesis. Our results indicate the promise of leveraging the learned program synthesizer to improve the dataset quality for programming by example tasks.

We consider extending our approach to synthesize more complicated real-world code as future work. For example, we will integrate our latent executor into large-scale pre-trained language models, which could further improve the performance of those program synthesis models taking natural language specifications. We will also study program synthesis problems with unbounded input ranges and different type signatures, which could be approached with the usage of subword tokenizers.

{%\small
\bibliographystyle{abbrv}
\bibliography{ref}
}

%\clearpage
%\input{checklist}
\clearpage
\appendix
\section{More Descriptions of the Karel Domain}
\label{app:karel-details}

We present the full grammar of the Karel language in Fig.~\ref{fig:karel-grammar}.

To represent the execution states, each Karel grid world has a maximum size of $18\times 18$, and each cell in the grid world is represented by a $16$-dimensional vector corresponding to the features in Table~\ref{tab:karel-state}. Therefore, each grid world is represented as a $16\times 18\times 18$ tensor.

\begin{figure}[h]
$$
\begin{array}{rcl}
\texttt{Prog p} & ::= & \texttt{def}~\texttt{run()}~\texttt{:}~\texttt{s} \\
\texttt{Stmt s} & ::= & \texttt{while(b)}~\texttt{:}~\texttt{s} \mid \texttt{repeat(r)}~\texttt{:}~\texttt{s} \mid~\texttt{$s_1$}~\texttt{;}~\texttt{$s_2$} \mid \texttt{a} \\
& | & \texttt{if(b)}~\texttt{:}~\texttt{s} \mid \texttt{ifelse(b)}~\texttt{:}~\texttt{$s_1$}~\texttt{else}~\texttt{:}~\texttt{$s_2$} \\
\texttt{Cond b} & ::= & \texttt{frontIsClear()} \mid \texttt{leftIsClear()} \mid \texttt{rightIsClear} \\
& | & \texttt{markersPresent()} \mid \texttt{noMarkersPresent()} \mid \texttt{not b} \\
\texttt{Action a} & ::= & \texttt{move()} \mid \texttt{turnRight()} \mid \texttt{turnLeft()} \\
& | & \texttt{pickMarker()} \mid \texttt{putMarker()} \\
\texttt{Cste r} & ::= & \texttt{0} \mid \texttt{1} \mid \texttt{...} \mid \texttt{19}
\end{array}
$$
\caption{Grammar for the Karel task.}
\label{fig:karel-grammar}
\end{figure}

\begin{table}[h]
    \centering
    \begin{tabular}{|c|}
        \hline
        Robot facing North \\
        \hline
        Robot facing East \\
        \hline
        Robot facing South \\
        \hline
        Robot facing West \\
        \hline
        Obstacle \\
        \hline
        Grid boundary \\
        \hline
        1 marker \\
        \hline
        2 markers \\
        \hline
        3 markers \\
        \hline
        4 markers \\
        \hline
        5 markers \\
        \hline
        6 markers \\
        \hline
        7 markers \\
        \hline
        8 markers \\
        \hline
        9 markers \\
        \hline
        10 markers \\
        \hline
    \end{tabular}
    \caption{Representation of each cell in the Karel state.}
    \label{tab:karel-state}
\end{table}

\section{Details in Model Architecture}
\label{app:model-architecture}
\subsection{Program Decoder}
\label{app:program-decoder}

Our model follows the encoder-decoder framework in prior work on neural program synthesis from input-output examples~\cite{devlin2017robustfill,bunel2018leveraging}, which includes an encoder for the input-output pairs, and a decoder to synthesize the program.

The program decoder is an LSTM (denoted as $\mathrm{LSTM}_D$), which decodes the program as a token sequence. Let $p_{t-1}$ be the decoded program token at step $t-1$, $E_p(p_{t-1})$ be the embedding vector of $p_{t-1}$, $h_{t-1}$ be the hidden state of the program decoder at step $t-1$, and $\hat I_{t-1}$ and $O$ be the sequences of vectors representing the input list elements and output list elements. We first compute attention vectors over both the input and output lists, following the double attention mechanism in RobustFill:

\[s_t^O=\mathrm{Attention}(h_{t-1}, O), \quad s_t^I=\mathrm{Attention}([h_{t-1}; s_t^O], \hat I_{t-1})\]

The notation $[a; b]$ means the concatenation of vectors $a$ and $b$. Then we calculate the output vector of the program decoder at step $t$ as $h_t=\mathrm{LSTM}_D(h_{t-1}, [E_p(p_{t-1}); s_t^I; s_t^O])$.

\paragraph{Input-Output Encoder.} For C program synthesis, our input-output encoder architecture is similar to RobustFill~\cite{devlin2017robustfill}. For each input-output pair, we use two bi-directional LSTMs~\cite{hochreiter1997long} to encode the input and output lists respectively. To capture the relationship between the input and output lists, the output list encoder computes attention vectors over the input list elements, using the standard attention mechanism~\cite{bahdanau2014neural,luong2015effective}. We also employ different encoder architectures for program synthesis tasks with other formats of input-output examples, as discussed in Sec.~\ref{sec:exp}.

%\yuandong{We probably want to briefly introduce ``double attention'' mechanism here, and why it is relevant in our setting?}

To capture the required arithmetic operation to convert from the program input to the output, we also include the output of the operation predictor $\hat{op}_t$ for program decoding, and we discuss the details later. Afterwards, the max pooling layer aggregates the representation of different IO pairs to generate a single vector:

\[m_t=\mathrm{MaxPool}_{j \in \{1,2,...,K\}}(\mathrm{tanh}(W[s_t^{I(j)}; s_t^{O(j)}; \hat{op}_t^{(j)}]))\]
Here the superscript $(j)$ indicates that the representation is for the $j$-th IO pair, and $W$ is a trainable weight matrix.
%\yuandong{We should mention why we have an operation encoder? The goal is to model the arithmetic structure of the input/ouput pairs. } \xinyun{Updated.}

To facilitate the prediction of long programs, we compute an attention vector over previously generated program tokens as follows:

\[d_t=\mathrm{Attention}(m_t, \{E_p(p_0), E_p(p_1), ..., E_p(p_{t-1})\})\]

Finally, the next token $p_t$ is sampled from $\mathbb{P}[p_t]=\mathrm{Softmax}(Vd_t)_{p_t}$ where $V$ is a trainable matrix.

%\label{sec:operation-predictor}

\subsection{Operation Predictor for Restricted C Domain}
\label{app:operation-predictor}
Training neural networks for mathematical reasoning is a challenging problem itself~\cite{saxton2019analysing,lample2019deep}, and jointly predicting the mathematical calculations and other program operations imposes extra burden on the program decoder. To mitigate the difficulty, we include a pre-computed table as part of the model input, which describes possible mathematical operations to derive an output value given the input number. For example, Fig.~\ref{fig:model-architecture}(d) shows that by applying the $O = 2 + I$ operation to the input $I=2$, the output $O=4$. For each valid input list value $C$, we include two operations $O = C + I$ and $O = C - I$ in the table. Then for each operation $O = C + I$, we enumerate all valid integer list values $I$, and we include the row $[O = C + I, I, O]$ in the table when $O$ is also within our bounded range. In this way, the table covers all possible integer addition and subtraction operations for valid input and output list values. \yuandong{Might want to say more by giving one concrete example.} \xinyun{Is it clearer in this way?}

With the pre-computed table, the operation predictor aims to predict the most possible program operation at the next step. First, we re-use the same embedding matrices as those in the input-output encoder, and compute the embedding vectors for each numerical element in the table. Let $R$ be the number of table rows. For the $i$-th row, we refer to the embedding vector of the input and output values as $r^{[i]}$ and $c^{[i]}$, respectively. Then we utilize $s_t^I$ and $s_t^O$ to compute the attention weights over the table columns of input and output values as follows:

\[wi^{[i]}_t=\mathrm{AttentionWeight}(s_t^I, \{r^{[i]}|i\in\{1, 2, ..., R\}\})\]

\[wo^{[i]}_t=\mathrm{AttentionWeight}(s_t^O, \{c^{[i]}|i\in\{1, 2, ..., R\}\})\]

Let $op^{[i]}$ be the operation in row $i$, then the probability of selecting the operation in the $i$-th row at step $t$ is

\[\mathbb{P}[op_t=op^{[i]}] \propto wi^{[i]}_t \cdot wo^{[i]}_t\]

Let $E_{op}(op)$ be the embedding vector of the operation $op$, then the operation predictor output is

\[\hat{op}_t=\sum_{i}\mathbb{P}[op_t=op^{[i]}]E_{op}(op^{[i]})\]

\yuandong{Is $E[op^{[i]}]$ the embedding? We might want to dedicate a table to list all the notations.} \xinyun{Yes this is the embedding. It might be better to show the important notations in figures, otherwise there might be too many notations to list.}

To train the operation predictor, we provide the training supervision at step $0$, when no transformation has been applied to the program input:
\begin{equation}
\mathcal{L}_{Op}=\mathrm{Loss}(wi_0^{[i]}, \mathbbm{1}[r^{[i]}=\hat I_0=I]) + \mathrm{Loss}(wo_0^{[i]}, \mathbbm{1}[c^{[i]}=O]) \label{eq:loss-op}
\end{equation}

\subsection{Latent Executor}
\label{app:latent-executor}

In RobustFill, the encoder only takes the initial input-output pairs as the input. On the other hand, in recent work on execution-guided program synthesis~\cite{chen2018execution,sun2018neural,Zohar2018AutomaticPSExtendExecution,Ellis2019WriteEAExtendExecution,odena2020bustle,nye2020representing}, the execution states of partial programs are leveraged as the model input to guide the subsequent program prediction. However, existing approaches mostly assume that the programs are sequential~\cite{Zohar2018AutomaticPSExtendExecution,Ellis2019WriteEAExtendExecution}, or require an interpreter of partial programs~\cite{chen2018execution}. To address these limitations, Nye et al. design neural networks to represent the partial program semantics when they are not well-defined~\cite{nye2020representing}. However, they need to train a separate neural module to represent each program operation, thus it is hard to scale beyond domain-specific languages.

In this work, we include another LSTM to approximate the program execution states, denoted as $\mathrm{LSTM}_E$. Let $\hat I_{t-1}$ be the input of $\mathrm{LSTM}_E$, which is the program input at step $t-1$. The output of $\mathrm{LSTM}_E$ is:

\[\mathrm{Exec}_t=\mathrm{LSTM}_E(h_t, \hat I_{t-1})\]

\subsubsection{Implementation for Restricted C Domain}
\label{app:latent-executor-c}

For our restricted C domain, the length of $\mathrm{Exec}_t$ is the same as $\hat I_{t-1}$, i.e., the input list length. Let $L$ be the length of input and output lists. Let $\mathbb{P}[I_t=v]$ be the probability that the execution result at step $t$ is $v$, then: \yuandong{How many possibilities $I_t$ would have? Exponential? $[-4,4]^5$ is like $9^5$?} \xinyun{Yes, the space is exponential. The idea is we assume that the probability of each element in the list is computed independently, since the output of $\mathrm{LSTM}_E$ encodes the context.} 

\[\mathbb{P}[I_{t,l}=v_l]=\mathrm{Softmax}(W_E\mathrm{Exec}_{t,l})_{v_l}\]

Here the subscript $l$ denotes that the representation is for the $l$-th list element, and $W_E$ is a trainable weight matrix.

\def\zz{\mathbb{Z}}

Finally, the approximated execution state $\hat{I}_t$ is the weighted sum of the embedding vectors of all possible program input integers $c \in [-4, 4] \cap \zz$ (where $\zz$ is the set of all integers): 

\[\hat I_{t,l}=\sum_{c \in [-4, 4]\cap \zz}\mathbb{P}[I_{t,l}=c]E_{io}(c)\]

Here $E_{io}(c)$ denotes the embedding vector of the list value $c$. At the next program decoding step, $\hat{I}_t$ will be fed into the encoder to replace the previous input list $\hat{I}_{t-1}$. 

\subsubsection{Implementation for Karel Domain}
\label{app:latent-executor-karel}

Similar to our restricted C domain, in our latent executor implementation for Karel domain, $\hat I_{t,l}$ is also the weighted sum of all possible execution states. As described in Appendix~\ref{app:karel-details}, each Karel state describes the following variables: (1) $(\text{robot}_X, \text{robot}_Y)$ denotes the position of the Karel robot, where $0 \leq \text{robot}_X, \text{robot}_Y < 18$; (2) $\text{robot}_{dir} \leq$ \{North, South, West, East\} denotes the robot orientation at $(\text{robot}_X, \text{robot}_Y)$; and (3) the number of markers in each grid. Therefore, we train 3 predictors on top of $\mathrm{LSTM}_E$ to predict these variables: (1) a trainable layer that outputs a ($18 \times 18$)-dimensional vector, representing the robot position; (2) a trainable layer that outputs a $4$-dimensional vector, representing the robot orientation; and (3) an LSTM that generates an 11-dimensional vector at each step, representing the number of markers in each grid. We apply the softmax to all output vectors to obtain the probability distributions of different variables.

Afterward, we combine the outputs of the predictors to construct a $16 \times 18 \times 18$-dimensional vector representing the Karel state, according to Table~\ref{tab:karel-state}, with the value of each dimension in $[0, 1]$. Note that Karel programs can not change the grid boundary and obstacles, thus we apply a mask on the predicted intermediate execution states to ensure that the features representing the grid boundary and obstacles remain the same, which are the last 2 dimensions described in Table~\ref{tab:karel-state}.

\section{Implementation Details}
\label{app:implementation-details}

All encoders and decoders in our models are 2-layer bi-directional LSTMs with the hidden size of 512. The embedding size is 1024. We use the Adam optimizer~\cite{kingma2014adam} for training. The learning rate starts from 1e-3, and is decayed by 0.9 for every 6000 timesteps. The batch size is 8. The training converges in 200K batch updates. The norm for gradient clipping is 5.0. All models are trained on a single GPU. The beam size is 64 for evaluating the model performance, and is 8 for iterative retraining due to the large size of the training set.

About the implementation of the Property Signatures~\cite{odena2020learning}, we further illustrate the key difference between our adaption for the restricted C domain and the original implementation in~\cite{odena2020learning} with the following example. Suppose an input-output pair is $([-4, 3, 1, 2, 1], [-4, 3, 3, 3, 3])$, when the feature is ``Input == Output?'', the corresponding property signature is ``False'' according to the implementation in~\cite{odena2020learning}, while the signature is ``[True, True, False, False, False]'' in our adapted implementation. Compared to the original implementation of property signatures, our adaptation better reveals which specific list elements are manipulated in the program. This modification makes our implementation of property signatures a much stronger baseline for the restricted C domain, because our C programs do not always perform the same manipulation steps over all elements in the input list, and sometimes change the values of only a subset of the input numbers.

\section{More Results of Iterative Retraining}
\label{app:retrain}

\begin{figure}
\begin{minipage}{0.25\textwidth}
\begin{minted}[fontsize=\scriptsize]{c}
I1: [2, 4, 1, 2, -3]
O1: [2, 4, 3, 2, -3]
I2: [1, 0, 1, -3, 4]
O2: [1, 0, 3, -3, 4]
I3: [2, 2, -4, 2, 0]
O3: [2, 2, 3, 2, 0]
I4: [0, -2, 3, 1, 3]
O4: [0, -2, 3, 1, 3]
I5: [-2, 1, 4, 0, 0]
O5: [-2, 1, 3, 0, 0]
\end{minted}
\end{minipage}
\begin{minipage}{0.40\textwidth}
\begin{minted}[fontsize=\scriptsize]{c}
int * func_1(int a[])
{
    int p_0 = 4;
    int l_7 = 2;
    int l_8 = 4;
    a[l_7] = 3;
    a[l_8] = a[p_0];
    return a;
}
\end{minted}
\end{minipage}
\begin{minipage}{0.33\textwidth}
\begin{minted}[fontsize=\scriptsize]{c}
int * func_1(int a[])
{
    int p_0 = 2;
    a[p_0] = 3;
    return a;
}
\end{minted}
\end{minipage}
\par\vspace{3mm}
\begin{minipage}{0.25\textwidth}
\begin{minted}[fontsize=\scriptsize]{c}
I1: [3, 1, 3, -2, -4]
O1: [3, 1, 2, -2, -4]
I2: [2, 0, -1, -1, 3]
O2: [2, 0, 2, -1, 3]
I3: [2, 0, -1, 4, 0]
O3: [2, 0, 2, 4, 0]
I4: [-2, -1, 3, 2, -4]
O4: [-2, -1, 2, 2, -4]
I5: [-4, 0, 3, 0, 1]
O5: [-4, 0, 2, 0, 1]
\end{minted}
\end{minipage}
\begin{minipage}{0.40\textwidth}
\begin{minted}[fontsize=\scriptsize]{c}
int * func_1(int a[])
{
    int p_0 = 2;
    int l_10 = 0;
    int l_1 = 4;
    l_10 = 2;
    for (p_0 = 2; p_0 >= 1; p_0--)
    {
        a[p_0] = 3;
        a[p_0] = 2;
        if (a[p_0])
            break;
        a[p_0] = a[l_1];
        a[p_0]++;
    }
    return a;
}
\end{minted}
\end{minipage}
\begin{minipage}{0.33\textwidth}
\begin{minted}[fontsize=\scriptsize]{c}
// Training on random programs
int * func_1(int a[])
{
    int p_0 = 2;
    int l_7 = 2;
    a[l_7] = 2;
    return a;
}

// After iterative retraining
int * func_1(int a[])
{
    int p_0 = 2;
    a[p_0] = 2;
    return a;
}
\end{minted}
\end{minipage}
\par\vspace{3mm}
\begin{minipage}{0.25\textwidth}
\begin{minted}[fontsize=\scriptsize]{c}
I1: [0, 4, 0, 4, 2]
O1: [0, 4, 0, 1, 1]
I2: [4, 0, 1, 1, 4]
O2: [4, 0, 1, 1, 1]
I3: [3, 2, 3, 0, 0]
O3: [3, 2, 3, 1, 1]
I4: [1, 1, 4, 0, 4]
O4: [1, 1, 4, 1, 1]
I5: [1, 3, 0, 1, 1]
O5: [1, 3, 0, 1, 1]
\end{minted}
\end{minipage}
\begin{minipage}{0.40\textwidth}
\begin{minted}[fontsize=\scriptsize]{c}
int * func_1(int a[])
{
    int p_0 = 0;
    int l_10 = 3;
    for (p_0 = 4; p_0 >= 0; p_0--)
    {
        a[p_0] = 3;
        a[p_0] = a[p_0];
        a[p_0] = 1;
        if (a[p_0])
            break;
    }
    a[l_10] = a[l_10];
    a[l_10] = a[p_0];
    return a;
}
\end{minted}
\end{minipage}
\begin{minipage}{0.33\textwidth}
\begin{minted}[fontsize=\scriptsize]{c}
int * func_1(int a[])
{
    int p_0 = 4;
    for (p_0 = 3; p_0 <= 4; p_0++)
    {
        a[p_0] = 1;
    }
    return a;
}
\end{minted}
\end{minipage}
\par\vspace{3mm}
\begin{minipage}{0.25\textwidth}
\begin{minted}[fontsize=\scriptsize]{c}
I1: [0, 3, -1, 0, 0]
O1: [4, 3, -1, 4, 4]
I2: [4, -3, 3, 4, 2]
O2: [4, -3, 3, 4, 4]
I3: [-4, 1, 0, 4, -2]
O3: [4, 1, 0, 4, 4]
I4: [0, 4, 3, 0, 4]
O4: [4, 4, 3, 4, 4]
I5: [2, 2, 0, 3, 2]
O5: [4, 2, 0, 4, 4]
\end{minted}
\end{minipage}
\begin{minipage}{0.40\textwidth}
\begin{minted}[fontsize=\scriptsize]{c}
int * func_1(int a[])
{
    int p_0 = 0;
    int l_11 = 3;
    for (p_0 = 2; p_0 >= 1; p_0--)
    {
        for (int p_1 = 4; p_1 >= 3; p_1--)
        {
            a[p_1] = 4;
        }
    }
    a[p_0] = a[l_11];
    return a;
}
\end{minted}
\end{minipage}
\begin{minipage}{0.33\textwidth}
\begin{minted}[fontsize=\scriptsize]{c}
int * func_1(int a[])
{
    int p_0 = 3;
    int l_7 = 0;
    a[l_7] = 4;
    for (p_0 = 4; p_0 >= 3; p_0--)
    {
        a[p_0] = 4;
    }
    return a;
}
\end{minted}
\end{minipage}
\par\vspace{3mm}
\begin{minipage}{0.25\textwidth}
\begin{minted}[fontsize=\scriptsize]{c}
I1: [1, 0, 0, 4, -3]
O1: [1, 1, 1, 4, -3]
I2: [-3, 0, 0, -2, 4]
O2: [1, 1, 1, -2, 4]
I3: [0, 2, -2, 4, -3]
O3: [1, 1, 1, 4, -3]
I4: [4, -2, 0, -2, 0]
O4: [1, 1, 1, -2, 0]
I5: [0, 2, -4, 2, 2]
O5: [1, 1, 1, 2, 2]
\end{minted}
\end{minipage}
\begin{minipage}{0.40\textwidth}
\begin{minted}[fontsize=\scriptsize]{c}
int * func_1(int a[])
{
    int p_0 = 4;
    for (p_0 = 1; p_0 >= 0; p_0--)
    {
        a[p_0] = 1;
        for (int p_1 = 2; p_1 >= 1; p_1--)
        {
            a[p_1] = 4;
            a[p_1] = a[p_0];
            if (a[p_1])
                break;
        }
    }
    return a;
}
\end{minted}
\end{minipage}
\begin{minipage}{0.33\textwidth}
\begin{minted}[fontsize=\scriptsize]{c}
int * func_1(int a[])
{
    int p_0 = 1;
    for (p_0 = 2; p_0 >= 0; p_0--)
    {
        a[p_0] = 1;
    }
    return a;
}
\end{minted}
\end{minipage}
\caption{\small More examples of predicted correct programs that are more concise than the randomly generated ground truth programs on C dataset. Left: input-output examples. Middle: the randomly generated ground truth program. Right: the predicted programs. Unless otherwise specified, the predicted programs come from the model trained on random programs.}
\label{fig:ex-retrain-c}
\vspace{-0.15in}
\end{figure}

\begin{figure}
\begin{minipage}{0.25\textwidth}
\begin{minted}[fontsize=\scriptsize]{c}
def run():
  repeat (5):
    ifelse (rightIsClear):
      move
    else:
      move
    putMarker
\end{minted}
\end{minipage}
\begin{minipage}{0.20\textwidth}
\begin{minted}[fontsize=\scriptsize]{c}
def run():
  repeat (5):
    move
    putMarker
\end{minted}
\end{minipage}
\begin{minipage}{0.25\textwidth}
\begin{minted}[fontsize=\scriptsize]{c}
def run():
  move
  turnRight
  ifelse (noMarkersPresent):
    repeat (2):
      putMarker
  else:
    pickMarker
  repeat (5):
    turnRight
\end{minted}
\end{minipage}
\begin{minipage}{0.25\textwidth}
\begin{minted}[fontsize=\scriptsize]{c}
def run():
  move
  turnLeft
  turnLeft
  ifelse (markersPresent):
    pickMarker
  else:
    putMarker
    putMarker
\end{minted}
\end{minipage}
\par\vspace{3mm}
\begin{minipage}{0.3\textwidth}
\begin{minted}[fontsize=\scriptsize]{c}
def run():
  pickMarker
  move
  ifelse (not rightIsClear):
    putMarker
    move
  else:
    move
    putMarker
    while (not rightIsClear):
      move
      putMarker
  putMarker
  turnRight
  move
\end{minted}
\end{minipage}
\begin{minipage}{0.20\textwidth}
\begin{minted}[fontsize=\scriptsize]{c}
def run():
  pickMarker
  move
  move
  putMarker
  putMarker
  turnRight
  move
\end{minted}
\end{minipage}
\begin{minipage}{0.20\textwidth}
\begin{minted}[fontsize=\scriptsize]{c}
def run():
  move
  turnRight
  repeat (5):
    pickMarker
  putMarker
\end{minted}
\end{minipage}
\begin{minipage}{0.25\textwidth}
\begin{minted}[fontsize=\scriptsize]{c}
def run():
  move
  repeat (4):
    pickMarker
  turnRight
\end{minted}
\end{minipage}
\par\vspace{3mm}
\begin{minipage}{0.3\textwidth}
\begin{minted}[fontsize=\scriptsize]{c}
def run():
  move
  ifelse (markersPresent):
    ifelse (frontIsClear):
      putMarker
    else:
      pickMarker
  else:
    while (rightIsClear):
      turnRight
  repeat (2):
    repeat (2):
      putMarker
    turnLeft
\end{minted}
\end{minipage}
\begin{minipage}{0.20\textwidth}
\begin{minted}[fontsize=\scriptsize]{c}
def run():
  move
  while (leftIsClear):
    turnLeft
  repeat (4):
    putMarker
\end{minted}
\end{minipage}
\begin{minipage}{0.25\textwidth}
\begin{minted}[fontsize=\scriptsize]{c}
def run():
  putMarker
  move
  ifelse (not leftIsClear):
    putMarker
  else:
    turnRight
  if (rightIsClear):
    pickMarker
\end{minted}
\end{minipage}
\begin{minipage}{0.20\textwidth}
\begin{minted}[fontsize=\scriptsize]{c}
def run():
  putMarker
  move
  putMarker
  if (rightIsClear):
    pickMarker
\end{minted}
\end{minipage}
\par\vspace{3mm}
\begin{minipage}{0.3\textwidth}
\begin{minted}[fontsize=\scriptsize]{c}
def run():
  while (not rightIsClear):
    while (noMarkersPresent):
      putMarker
    move
    turnLeft
    ifelse (rightIsClear):
      while (noMarkersPresent):
        putMarker
        turnLeft
        turnLeft
        move
      turnLeft
      move
    else:
      turnLeft
  turnLeft
  turnLeft
  repeat (4):
    move
  turnLeft
\end{minted}
\end{minipage}
\begin{minipage}{0.25\textwidth}
\begin{minted}[fontsize=\scriptsize]{c}
def run():
  putMarker
  move
  if (noMarkersPresent):
    putMarker
    turnRight
    move
    putMarker
    turnRight
  while (not frontIsClear):
    turnRight
    move
    turnRight
  repeat (3):
    move
  turnLeft
\end{minted}
\end{minipage}
\begin{minipage}{0.25\textwidth}
\begin{minted}[fontsize=\scriptsize]{c}
def run():
  repeat (6):
    if (markersPresent):
      repeat (9):
        pickMarker
    putMarker
    move
  putMarker
  turnRight
\end{minted}
\end{minipage}
\begin{minipage}{0.15\textwidth}
\begin{minted}[fontsize=\scriptsize]{c}
def run():
  repeat (6):
    putMarker
    move
  putMarker
  turnRight
\end{minted}
\end{minipage}
\par\vspace{3mm}
\begin{minipage}{0.3\textwidth}
\begin{minted}[fontsize=\scriptsize]{c}
def run():
  turnRight
  move
  turnRight
  move
  while (not rightIsClear):
    move
  pickMarker
  ifelse (not leftIsClear):
    move
  else:
    move
\end{minted}
\end{minipage}
\begin{minipage}{0.15\textwidth}
\begin{minted}[fontsize=\scriptsize]{c}
def run():
  turnRight
  move
  turnRight
  move
  pickMarker
  move
\end{minted}
\end{minipage}
\begin{minipage}{0.30\textwidth}
\begin{minted}[fontsize=\scriptsize]{c}
def run():
  while (frontIsClear):
    ifelse (markersPresent):
      move
    else:
      putMarker
\end{minted}
\end{minipage}
\begin{minipage}{0.20\textwidth}
\begin{minted}[fontsize=\scriptsize]{c}
def run():
  while (frontIsClear):
    putMarker
    move
\end{minted}
\end{minipage}
\caption{\small Examples of predicted correct programs that are more concise than the randomly generated ground truth programs on Karel dataset. 1st and 3rd columns: the randomly generated ground truth programs. 2nd and 4th: the corresponding predicted programs. The predictions come from the model trained on random programs.}
\label{fig:ex-retrain-karel}
\vspace{-0.15in}
\end{figure}

Figure~\ref{fig:ex-retrain-c} presents more examples of predicted correct programs that are more concise than the randomly generated ground truth programs on C dataset.

Figure~\ref{fig:ex-retrain-karel} presents more examples of predicted correct programs that are more concise than the randomly generated ground truth programs on Karel dataset. Note that the predicted Karel program is not semantically equivalent to the annotated ground truth in many cases. The main reason is because the randomly generated ground truth program might include redundant branching statements, i.e., the conditions always evaluate to true or false for all program inputs in the specification and the held-out test cases.

We present the numerical results of iterative retraining on Karel and C benchmarks in Table~\ref{tab:karel-iterative-retraining} and Table~\ref{tab:c-iterative-retraining} respectively.

\begin{table}[t]
\caption{Results of iterative retraining on Karel dataset.}
\label{tab:karel-iterative-retraining}
\centering
\begin{tabular}{lc|ccccc}
\toprule
\textbf{Iters} & \textbf{100\%}  & \textbf{10\%} & \textbf{20\%} & \textbf{30\%} & \textbf{40\%} & \textbf{50\%}\\
\hline
\multicolumn{7}{c}{Generalization Accuracy} \\
\hline
1 & 86.04\% & 70.92\% & 75.16\% & 78.84\% & 80.88\% & 82.08\% \\
2 & 89.28\% & 76.20\% & 78.40\% & 81.08\% & 82.40\% & 83.40\% \\
3 & 89.36\% & 78.12\% & 81.20\% & 83.68\% & 84.24\% & 86.32\% \\
\hline
\multicolumn{7}{c}{Exact Match Accuracy} \\
\hline
1 & 39.40\% & 36.20\% & 37.20\% & 38.36\% & 40.20\% & 40.04\% \\
2 & 41.56\% & 37.24\% & 37.28\% & 39.24\% & 39.72\% & 39.16\% \\
3 & 41.16\% & 36.56\% & 38.16\% & 38.68\% & 38.72\% & 39.64\% \\
\bottomrule
\end{tabular}
\end{table}

\begin{table}[t]
\caption{Results of iterative retraining on C dataset.}
\label{tab:c-iterative-retraining}
\centering
\begin{tabular}{lc|ccccc}
\toprule
\textbf{Iters} & \textbf{100\%}  & \textbf{10\%} & \textbf{20\%} & \textbf{30\%} & \textbf{40\%} & \textbf{50\%}\\
\midrule
1 & 55.2\% & 11.9\% & 26.4\% & 39.1\% & 45.2\% & 48.5\% \\
2 & 56.0\% & 39.6\% & 43.9\% & 48.7\% & 51.9\% & 54.1\% \\
3 & 56.5\% & 41.7\% & 44.4\% & 49.4\% & 52.8\% & 54.4\% \\
\bottomrule
\end{tabular}
\end{table}
\end{document}